\documentclass[times,mathptm,usenatbib,longnamesfirst]{mn2e}
\RequirePackage{graphicx}
\begin{document}
\title[A SCUBA survey of Orion]{A SCUBA survey of Orion -- the low-mass end of the core mass function}
\author[Nutter \& Ward-Thompson]
{D. Nutter\thanks{E-mail: David.Nutter@astro.cf.ac.uk}, 
D. Ward-Thompson\\
Department of Physics and Astronomy, Cardiff University, Queens Buildings, Cardiff, CF24 3AA}

\maketitle

\begin{abstract}
We have re-analysed all of the SCUBA archive data of the Orion star-forming regions. We have put together all of the data taken at different times by different groups. Consequently we have constructed the deepest submillimetre maps of these regions ever made. There are four regions that have been mapped: Orion A North \& South, and Orion B North \& South. We find that two of the regions, Orion A North and Orion B North, have deeper sensitivity and completeness limits, and contain a larger number of sources, so we concentrate on these two. We compare the data with archive data from the Spitzer Space Telescope to determine whether or not a core detected in the submillimetre is pre-stellar in nature. We extract all of the pre-stellar cores from the data and make a histogram of the core masses. This can be compared to the stellar initial mass function (IMF). We find the high-mass core mass function follows a roughly Salpeter-like slope, just like the IMF, as seen in previous work. Our deeper maps allow us to see that the core mass function (CMF) turns over at $\sim 1.3 M_\odot$, about a factor of 4 higher than our completeness limit. This turnover has never previously been observed, and is only visible here due to our much deeper maps. It mimics the turnover seen in the stellar IMF at $\sim 0.1M_\odot$. The low-mass side of the CMF is a power-law with an exponent of $0.35 \pm 0.2$, which is consistent with the low-mass slope of the young cluster IMF of $0.3 \pm 0.1$. This shows that the CMF continues to mimic the shape of the IMF all the way down to the lower completeness limit of these data at $\sim$0.3M$_\odot$.
\end{abstract}


\section{Introduction}\label{intro}
The earliest stages of low-mass star formation are now reasonably well understood. Stars are known to form within molecular clouds, or more precisely, within prestellar cores \citep{1994MNRAS.268..276W,1996A&A...314..625A,1999MNRAS.305..143W}, which are the density peaks of the molecular cloud that have become gravitationally bound. The youngest stars that have been observed are the Class 0 protostars \citep{1993ApJ...406..122A,2001A&A...369..155C}, which are still accreting material from their surrounding envelope. This envelope is the remnant of the prestellar core within which the protostar was born. For a detailed review of this process, see \citet{2000prpl.conf...59A,dwt_ppv}.

However, the formation mechanism for the lowest mass stars and brown dwarfs is less well understood, partially due to the observational difficulties in imaging these low-mass objects. Given the assumption that the minimum mass of a star that forms by the gravitational fragmentation of a molecular cloud is the Jeans mass \citep{1992MNRAS.256..641L}, it has been suggested that these very low-mass objects form by a different mechanism to solar-mass stars. There have been a number of mechanisms suggested, including star formation in low-mass cores that are created in shock compressed layers \citep[e.g.][]{2005A&A...430.1059B}, or ejection from a more massive core before it has time to accrete very much mass \citep{2001AJ....122..432R,2002MNRAS.332L..65B,2002MNRAS.336..705B,2004A&A...414..633G,2004A&A...419..543G,2004A&A...423..169G}. These mechanisms and others are reviewed in \citet{ant_ppv}.

A recent important observational result in the study of star formation is the discovery that the mass spectrum of prestellar cores mimics the Salpeter-like power-law slope of the stellar initial mass function at high masses \citep{1998A&A...336..150M}.
However, due to completeness limits, neither this nor more recent studies have probed the relation between the core mass function (CMF) and the stellar IMF at low masses. Knowledge of the relation between the CMF and the IMF in this mass regime is required to understand the formation mechanism of very low-mass stars and brown dwarfs.

\begin{figure}
\includegraphics[angle=0,width=83mm]{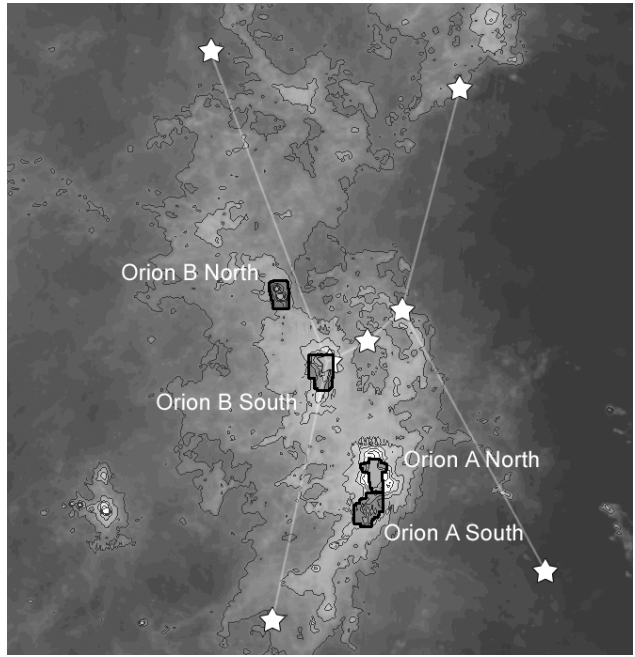}
\caption{IRAS 100~$\mu$m map of the Orion molecular cloud \citep{IRAS}, as well as the familiar stars of the constellation. The north-south extent of the image is 20 deg, which at an assumed distance of 400 pc, corresponds to 140 pc. The extents of the four regions that have been mapped in the submillimetre are outlined in black. Our names for these regions are labelled.}
\label{orion_overview}
\end{figure}

The Orion molecular cloud is the closest high mass star-forming region \citep[][and references therein]{1989ARA&A..27...41G}. The distances to the front and rear edges of the Orion molecular cloud were determined by \citet{1994A&A...289..101B} to be $\sim 320$ pc and $\sim 500$ pc respectively. We therefore assume a canonical distance to the Orion star-forming regions of 400 pc.
The molecular cloud is composed of two parts. The Orion A cloud (L1640/1641) is located in the southern half of the Orion constellation. It contains OMC1, which is the most active site of current star formation in the region, and houses the Trapezium cluster of O and B stars. The Orion B cloud (L1630) lies approximately 5 degrees (35 pc) north of Orion A and is made up of two sub-clouds. The Orion B South cloud contains the star-forming region NGC 2023/2024, and is the formation site of the only O stars in Orion B. This part of the cloud is the location of B33, the famous Horsehead nebula \citep[see][and references therein]{2006MNRAS.369.1201W}. Orion B North is located 2.5 degrees (17 pc) north of NGC 2024, and contains the NGC 2068/2071 and HH24-26 star-forming regions. The Orion molecular cloud is illustrated in Figure~\ref{orion_overview} which shows the IRAS 100~$\mu$m data \citep{IRAS} produced using the {\em skyview} interface \citep{1994ASPC...61...34M}, along with the regions that have been mapped in the submillimetre.

In this study, we have combined all of the SCUBA wide-field scan-map data taken to date of the Orion molecular cloud, to produce the deepest and widest-area submillimetre maps of this cloud. With these data, we investigate the low-mass end of the CMF, and also compare the mass spectra for the different regions of the molecular cloud in a self consistent way. We further provide a catalogue of the size and mass of each of the detected cores within the Orion molecular cloud.

\section{Observations}\label{observations}
The submillimetre data presented in this study were obtained using the Submillimetre Common User Bolometer Array (SCUBA) on the James Clerk Maxwell Telescope (JCMT). This instrument takes observations at 450 and 850~$\mu$m simultaneously through the use of a dichroic beam-splitter. The telescope has a resolution of 8 arcsec at 450$~\mu$m and 14 arcsec at 850~$\mu$m. The data presented here were acquired from the JCMT data archive, operated by the Canadian Astronomy Data Centre. Sub-sets of these data have been published previously \citep{1999ApJ...510L..49J,JohnstoneOriAS,2001ApJ...556..215M,2001A&A...372L..41M,2001ApJ...559..307J,2006ApJ...639..259J,2006MNRAS.369.1201W}, though all the data have been re-reduced using a consistent method for this study.

The observations were carried out over 30 separate nights between February 1998 and March 2002 using the scan-map observing mode. A scan-map is made by scanning the array across the sky. The scan direction is 15.5$^\circ$ from the axis of the array in order to achieve Nyquist sampling. The array is rastered across the sky to build up a map several arcminutes in extent.

Time-dependent variations in the sky emission were removed by chopping the secondary mirror at 7.8 Hz. Due to a scan-map being larger in size than the chop throw, each source in the map appears as a positive and a negative feature. In order to remove this dual-beam function, each region is mapped six times, using chop throws of 30, 44 and 68 arcsec in both RA and Dec \citep{1995mfsr.conf..309E}. The dual-beam function is removed from each map in Fourier space by dividing each map by the Fourier transform of the dual-beam function, which is a sinusoid. The multiple chop-throws allow for cleaner removal of the dual beam function in Fourier space. The maps are then combined, weighting each map to minimise the noise introduced at the spatial frequencies that correspond to zeroes in the sinusoids. Finally the map is inverse Fourier transformed, at which point it no longer contains the negative sources \citep{SURF}.

\begin{figure*}
\begin{minipage}[t]{80mm}
\includegraphics[angle=0,width=80mm]{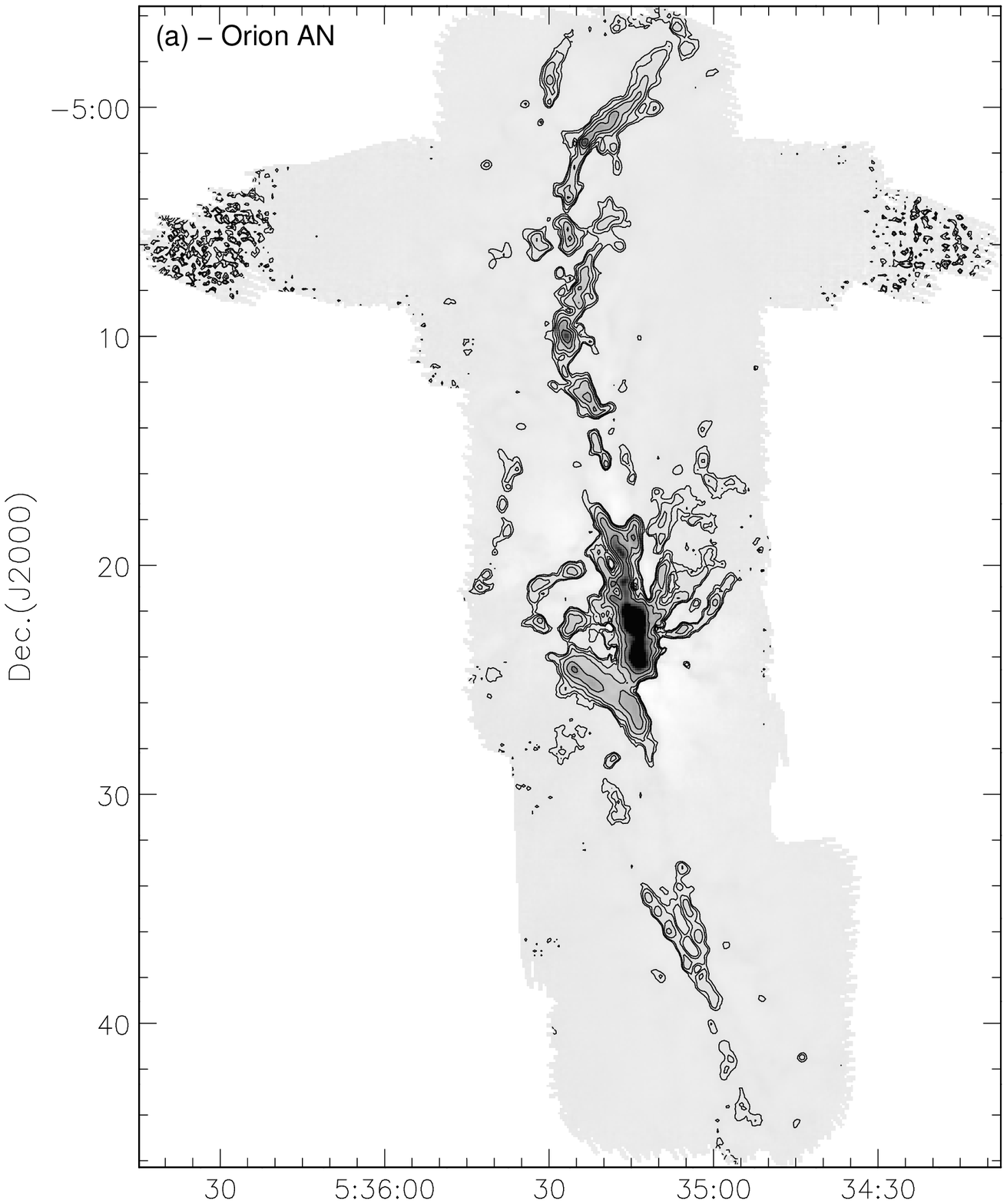} 
\end{minipage}
\begin{minipage}[t]{80mm}
\includegraphics[angle=0,width=80mm]{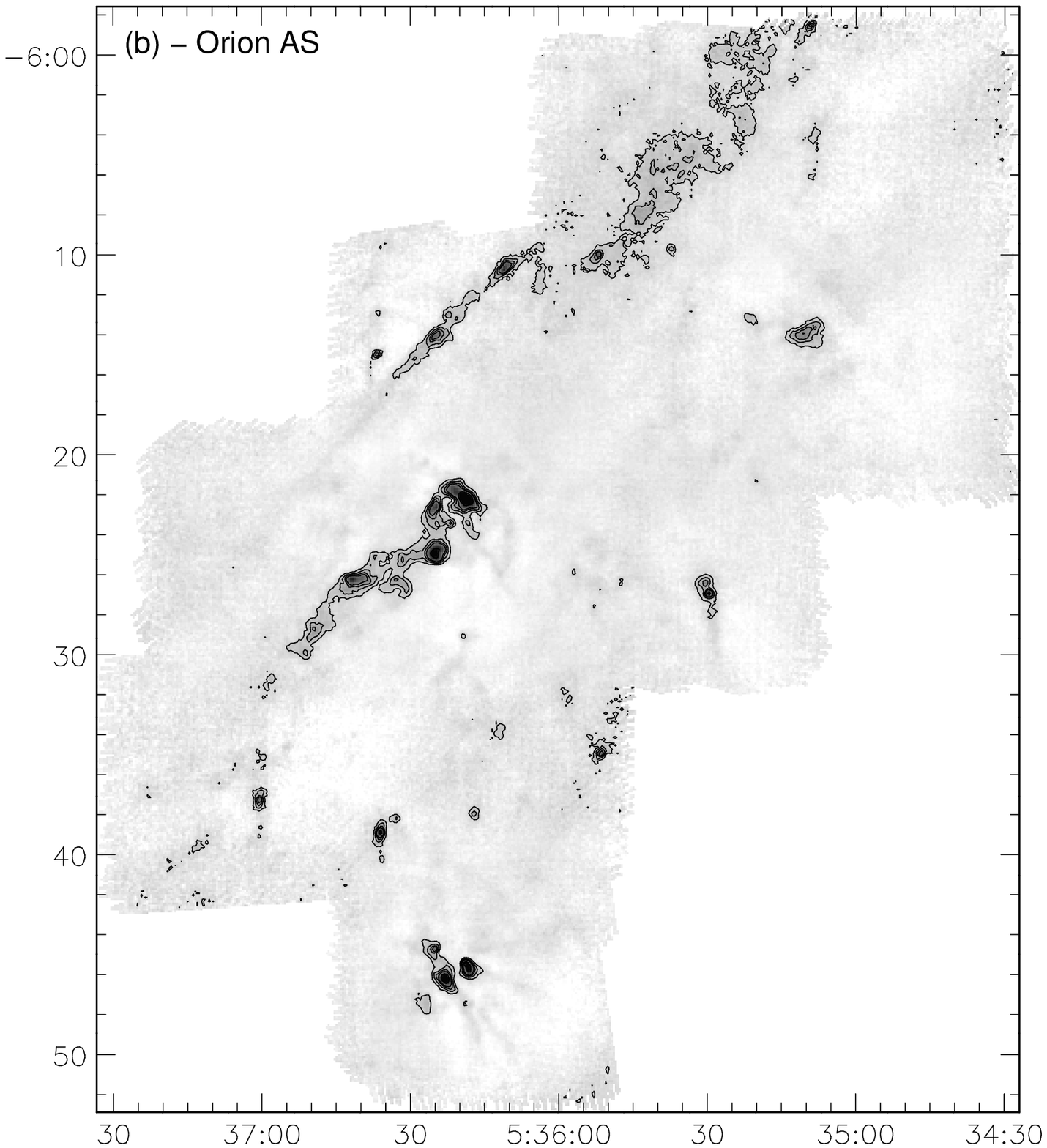}
\end{minipage}
\begin{minipage}[t]{80mm}
\vspace{0pt}
\includegraphics[angle=0,width=80mm]{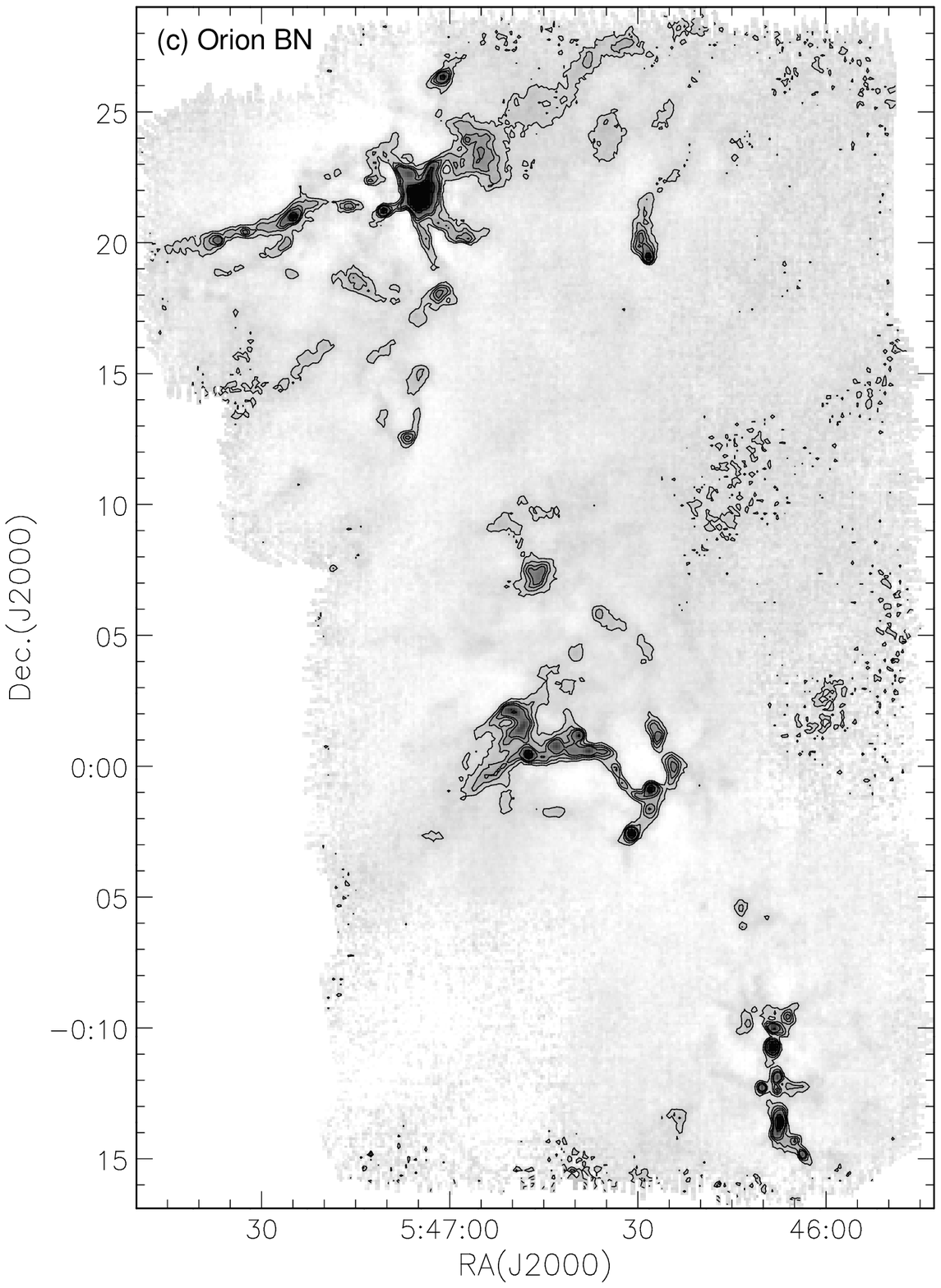}
\end{minipage}
\begin{minipage}[t]{80mm}
\vspace{0pt}
\includegraphics[angle=0,width=80mm]{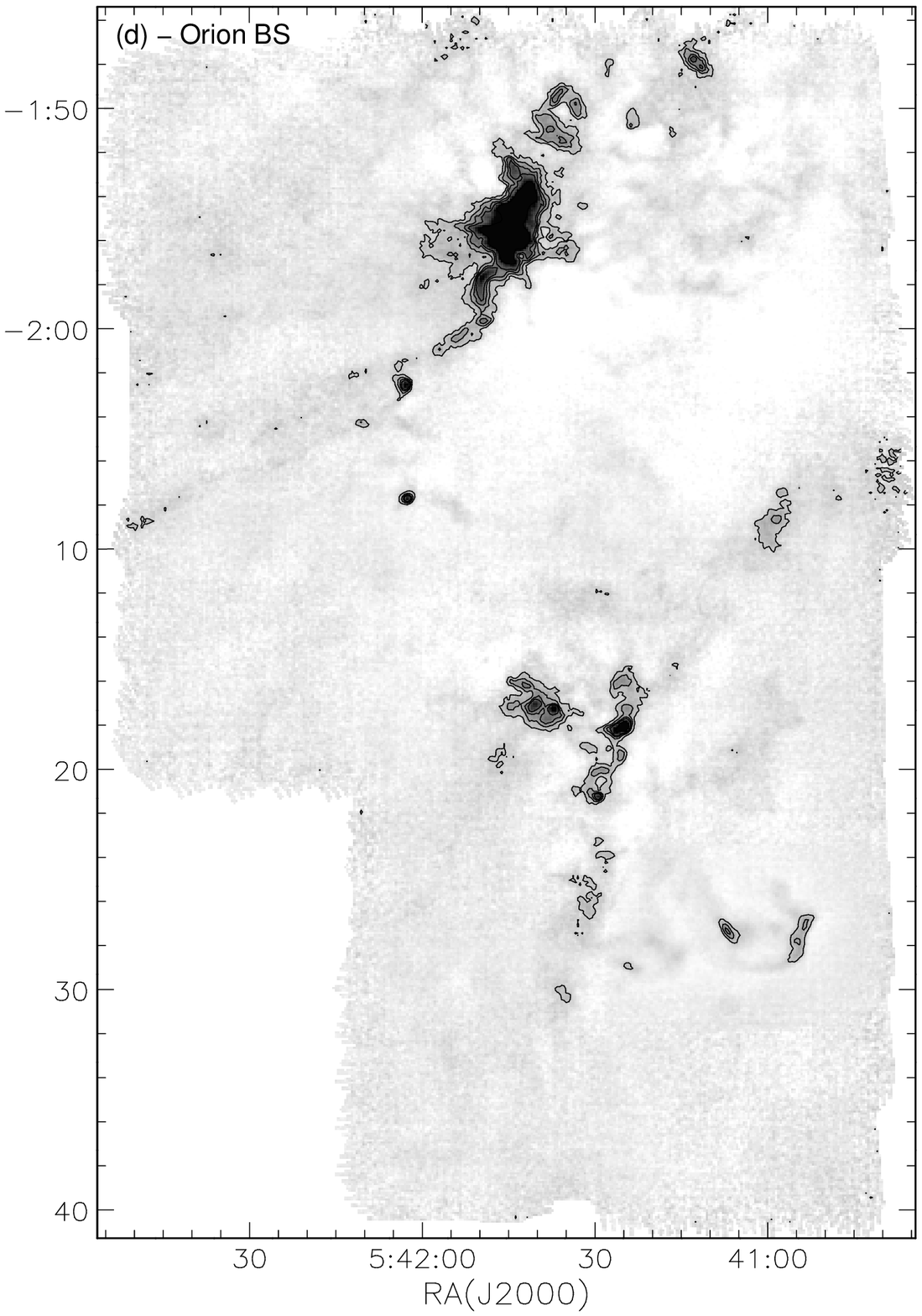}
\end{minipage}
\caption{The four maps of the Orion molecular cloud at 850~$\mu$m. 
a) Orion AN, with contour levels at 5, 10, 20, 40, 80 and 160$\sigma$, where $\sigma$ is the mean noise level over the whole of the map, equal to 19 mJy/beam. 
b) Orion AS, with contours at 5, 10, 15, 20 and 40$\sigma$, where $\sigma$ is equal to 18 mJy/beam. 
c) Orion BN, with contours at 5, 10, 15, 20 and 40$\sigma$, where $\sigma$ is equal to 16 mJy/beam. 
d) Orion BS, with contours at 5, 10, 15, 20 and 40$\sigma$, where $\sigma$ is equal to 23 mJy/beam.} 
\label{Orion_maps}
\end{figure*}

The submillimetre zenith opacity at 450 and 850$~\mu$m was determined using the `skydip' method and by comparison with polynomial fits to the 1.3~mm sky opacity data, measured at the Caltech Submillimeter Observatory \citep{2002MNRAS.336....1A}. The sky opacity at 850$~\mu$m varied from 0.12 to 0.51, with a median value of 0.26. These correspond to a 450$~\mu$m opacity range of 0.5 to 3.1, and a median value of 1.4. 

The data were reduced in the normal way using the SCUBA User Reduction Facility \citep{SURF}. Noisy bolometers were removed by eye, and the baselines, caused by chopping onto sky with a different level of emission, were removed using the {\em \small MEDIAN} filter. Calibration was performed using observations of the planets Uranus and Mars, and the secondary calibrators HLTau and CRL618 \citep{1994MNRAS.271...75S} taken during each shift. We estimate that the absolute calibration uncertainty is $\pm7\%$ at 850$~\mu$m and $\pm15\%$ at 450$\mu$m, based on the consistency and reproducibility of the calibration from map to map. 

The submillimetre data make up 4 separate maps. The relative positions of these maps are illustrated in Figure \ref{orion_overview}. The most northerly map of the four is centred on the northern star-forming complex in Orion B, including NGC 2071, NGC 2068 and HH24/25/26. Hereafter, we refer to this map as Orion BN. The Orion B South map (hereafter Orion BS) encompasses the southern star-forming component of Orion B, and includes NGC 2023, NGC 2024 and the Horsehead nebula (B33). The Orion A cloud is also made up of two maps, the northernmost of which (Orion A North -- hereafter Orion AN) includes OMC 1, 2 and 3. Directly to the south of Orion AN is Orion A South (hereafter Orion AS), which follows the Orion A filament to the south-east. The area covered by each of the maps is given in Table~\ref{noise_levels}.

The 850~$\mu$m data for the four regions are shown in Figures \ref{Orion_maps}(a-d). These figures show that the data contain a large number of dense cores, which are usually associated with larger scale filamentary structure. The main difference in the general morphology of the different regions is that they vary in richness, for example the cores in the Orion AN map are more numerous and more closely packed than in the other three regions. 

\section{Results}\label{results}
\subsection{Submillimetre continuum data}
In order to produce the deepest maps possible, the data are made up from a number of different observing programs, therefore the map coverage is not uniform for each region. In addition, the weather conditions varied when the data were taken. As a result, the noise levels vary across each map. In order to quantify the variation of the noise level across the map, a noise-map was generated using the following technique. 

The bright sources in the map were masked in order to prevent the rapid change in flux density at the positions of these sources from contaminating the noise estimate. For the same reason, the large scale structure of the map was removed by smoothing the map and subtracting this from the data. The noise-map was then built up by measuring the standard deviation of the resulting data at each point in an aperture of radius 50 arcsec. The gaps in the noise-map caused by the masking of the bright sources were filled in by interpolation from the surrounding regions. 

\begin{figure}
\includegraphics[angle=0,width=83mm]{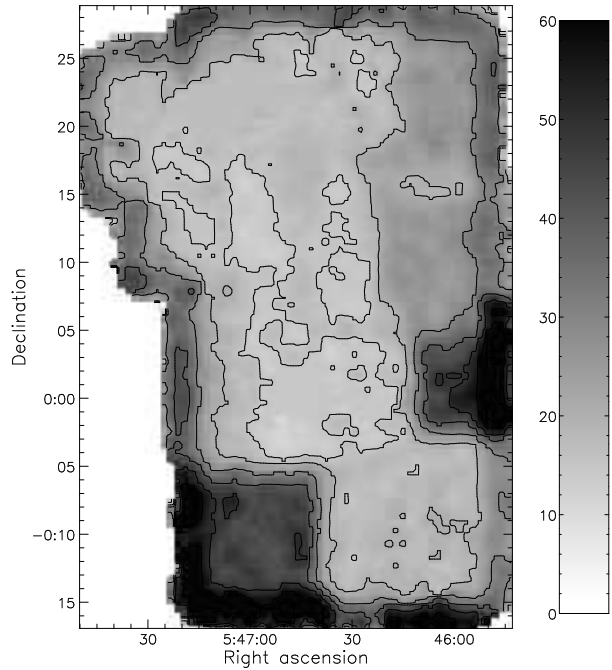}
\caption{The 850~$\mu$m noise-map for the Orion BN region. Contours are 10, 15, 20, 30, 40, 50, 60 \& 70 mJy/beam. See text for details.}
\label{Orion BN_noisemap}
\end{figure}

To illustrate this, the noise-map for the Orion BN is shown in Figure \ref{Orion BN_noisemap}, where the grey scale indicates the noise level in the map. As can clearly be seen, the noise level is lower towards the centre of the map due to increased coverage by the different observing programs. This effect is seen in all of the different regions. Table~\ref{noise_levels} gives the average noise levels, measured using the noise-maps. The quoted value corresponds to the average noise level measured over all parts of the map which contain detected sources. For comparison, previous authors using different subsets of the SCUBA data of Orion have found values for the 1$\sigma$ 850~$\mu$m noise level in the range $22-40$ mJy/beam \citep{1999ApJ...510L..49J,JohnstoneOriAS,2001A&A...372L..41M,2001ApJ...559..307J,2006ApJ...639..259J}.

\begin{table}
\begin{center}
\caption{The statistics for each of the four mapped regions. The number of cores detected in each region is given in column 2. The average 1$\sigma$ noise levels for the 850~$\mu$m and 450~$\mu$m data are given in columns 3 and 4 respectively. The 450~$\mu$m noise level is based on the data after smoothing to the 850 $\mu$m resolution.}
\label{noise_levels}
\vspace{0.5cm}
\begin{tabular}{ccccc} \hline
Cloud 		& Map		& No. of	& \multicolumn{2}{c|}{1$\sigma$ noise level} 	\\
Name 		& Area		& cores		& 850~$\mu$m	& 450~$\mu$m		\\ 
		& (sqdeg)	&		& (mJy/beam)	& (mJy/beam)		\\ \hline
Orion AN	& 0.23		& 172		& 19		& 67			\\
Orion AS	& 0.41		& 62		& 18		& 60			\\ 
Orion BN	& 0.32		& 100		& 16		& 39			\\
Orion BS	& 0.45		& 59		& 23		& 68			\\ \hline
\end{tabular}
\end{center}
\end{table}

In order to extract the properties of each core in a consistent manner across all of the maps, each map was divided by the noise-map described above, to produce a signal to noise (S/N) map. The significance and dimensions of sources were determined using this S/N map using the following criteria. 

A source was deemed real if it had a peak flux density $> 5\sigma$ relative to the local background. Close peaks were separated into two sources if the difference between the peak of the weaker source and the minimum between the sources was $> 3\sigma$. Elliptical apertures were placed on each source and approximately matched to the position of the $3\sigma$ contour. The flux density for each source was then determined by placing the aperture, defined using the S/N map, on the original data. In each case the flux density was measured relative to the local background level by placing a `sky' aperture on a nearby blank area of the map. The number of cores detected in each region is given in Table~\ref{noise_levels}, and the properties of each core are listed in Table~\ref{cores_table} in Appendix~A. 

\begin{figure*}
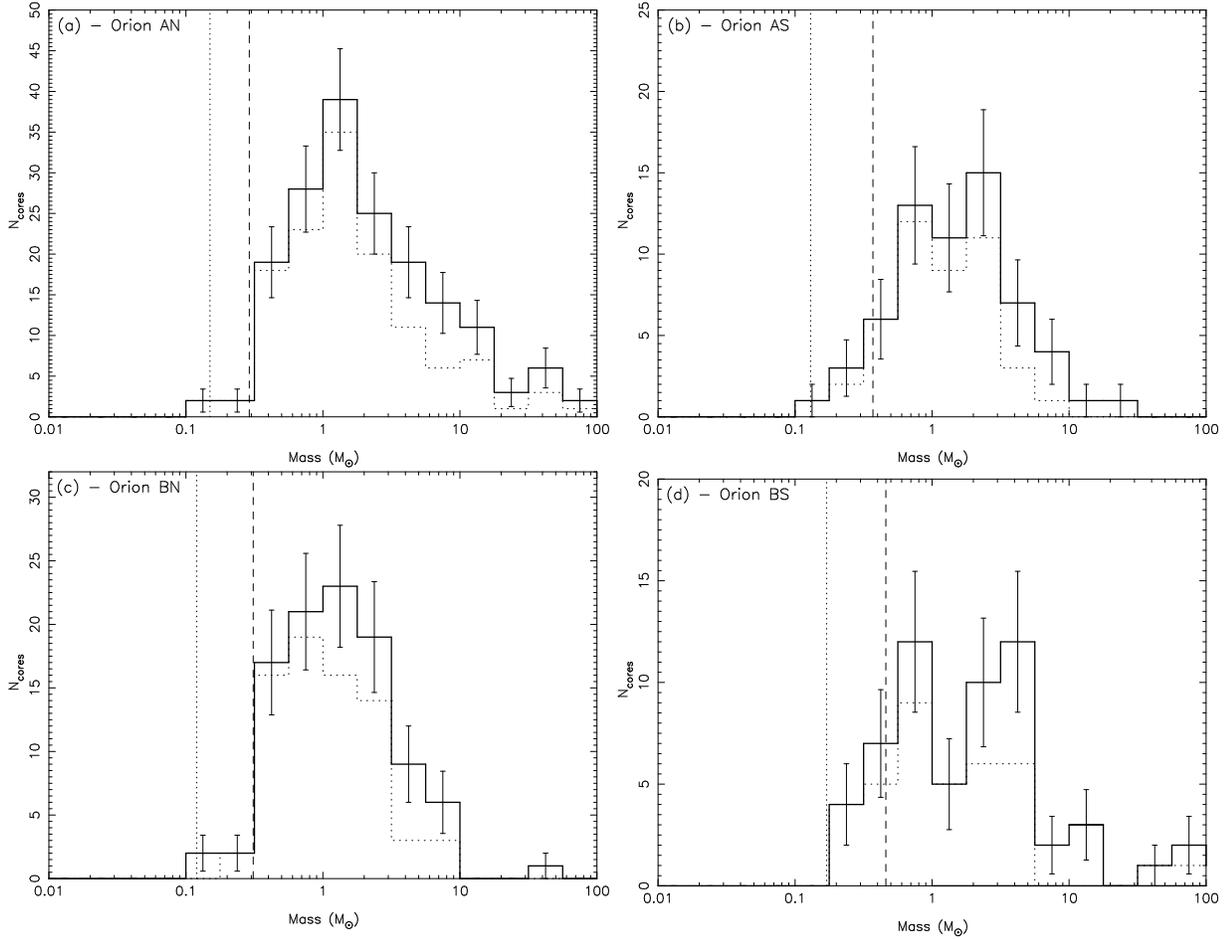

\begin{minipage}[t]{80mm}
\includegraphics[angle=270,width=80mm]{./fig4a.ps} 
\end{minipage}
\begin{minipage}[t]{80mm}
\includegraphics[angle=270,width=80mm]{./fig4b.ps}
\end{minipage}
\begin{minipage}[t]{80mm}
\vspace{0pt}
\includegraphics[angle=270,width=80mm]{./fig4c.ps}
\end{minipage}
\begin{minipage}[t]{80mm}
\vspace{0pt}
\includegraphics[angle=270,width=80mm]{./fig4d.ps}
\end{minipage}
\caption{The core mass functions for the four mapped regions. The solid histogram includes all cores that were detected at 850 $\mu$m. The dotted histogram shows only those cores which do not contain a Class I protostar, as determined using mid-IR Spitzer data. The error bars are the statistical error due to the number of cores in each bin. The dotted and dashed vertical lines show the 5$\sigma$ sensitivity and completeness limit respectively for each region. All detected submillimetre sources are included in these mass functions.}
\label{Orion_CMFs}
\end{figure*}

This strategy for characterising individual cores was applied to each source in the data with the exception of those sources that are embedded in a large amount of extended emission and are very close to neighbouring sources. In the case of these sources, their proximity to their neighbours means that the aperture size is necessarily reduced relative to the $3\sigma$ defined aperture discussed above. In order to compensate for this effect, the surrounding extended emission above the $3\sigma$ contour was partitioned between the cores relative to their peak flux density. 

Due to the poorer atmospheric opacity at 450~$\mu$m, the sources in the 450~$\mu$m data have significantly lower signal to noise ratios compared to the 850~$\mu$m data. The 450~$\mu$m data were therefore smoothed such that the FWHM resolution matched that of the 850~$\mu$m data. The sources were then extracted from the 450~$\mu$m data in a similar manner to that described above, with a signal to noise map being generated for each region, and a detection criterion of 5$\sigma$ applied. Each source that was detected at 450~$\mu$m has a corresponding detection at 850~$\mu$m, though not all sources that were detected at 850~$\mu$m were detected at 450~$\mu$m. The integrated flux was measured for the sources that were detected at 450~$\mu$m using apertures with the same position and dimensions as those used for the 850~$\mu$m sources. This is to allow a direct comparison of the data at the different wavelengths. The 450~$\mu$m properties for each core are given in Table~\ref{cores_table} in Appendix~A.

\subsection{Spitzer IRAC results}
Mid-IR data for the four regions taken using the IRAC camera \citep{2004ApJS..154...10F} on board the Spitzer space telescope were retrieved from the Spitzer data archive. These data have wavelengths of 3.6, 4.5, 5.8 and 8.0~$\mu$m. The data were calibrated using the IRAC science BCD pipeline prior to retrieval from the archive \citep{IracHandbook}. 

Each waveband of the IRAC data was independently searched for point sources that lie within the 3$\sigma$ contour of the submillimetre detected cores. The flux density of each source was measured in a circular aperture, the position and size of which was fixed for all wavelengths. Extended emission from the surrounding cloud was removed using an off-source `sky' aperture. The magnitudes of the sources were calculated using the zero magnitude flux densities given in the Spitzer Observer's Manual \citep{SpitzerUserGuide}.

Sources that were detected in all 4 IRAC bands were classified as Class I or Class II YSOs or as stars according to their [5.8] - [8.0] and [3.6] - [4.5] colours, as described in \citet{2004ApJS..154..363A}. Sources that were detected in either 2 or 3 IRAC bands, were classified according to their spectral index \citep{1987IAUS..115....1L}. Submillimetre sources that contain a Class I YSO were removed from our sample to remove protostellar contamination of the pre-stellar core statistics. 

\section{Core Mass Functions}

The masses of the cores are calculated from their 850~$\mu$m integrated flux density, which is assumed to be optically thin. The core mass is calculated using the following:
\begin{equation}
M=\frac{S_{850}D^2}{\kappa_{850}B_{850,T}},
\end{equation}
\noindent
where $S_{850}$ is the 850~$\mu$m flux density, $D$ is the distance to the source, $\kappa_{850}$ is the mass opacity of the gas and dust, and $B_{850,T}$ is the Planck function at temperature $T$. We assume a distance to the region of 400 pc \citep{1994A&A...289..101B}, a temperature of 20~K for the Orion cores \citep{1996A&A...312..569L,2001ApJ...556..215M,2006ApJ...639..259J}, and a mass opacity of ${\rm 0.01~cm^2g^{-1}}$ \citep[see][for a detailed discussion of both this value of $\kappa_{850}$ in particular and of this method of obtaining masses in general]{1993ApJ...406..122A,1996A&A...314..625A,1999MNRAS.305..143W}. We note that the assumption of a single temperature for all of the cores could potentially introduce a systematic bias in the calculated mass if there is a correlation between the temperature and mass of the cores. 
 
The core mass functions for the four regions are shown in Figures \ref{Orion_CMFs}(a-d). The error bars shown are the $\sqrt{N}$ counting uncertainty due to the number of cores in each mass bin. The dotted vertical lines show the $5\sigma$ sensitivity of the map. This is based on the average noise level as shown in Table~\ref{noise_levels}, and converted to a mass using the assumptions of distance, temperature and dust opacity given above. This signifies the minimum detectable point source in the map. The dashed lines show the $5\sigma$ completeness limit of the maps. The completeness limit is higher than the sensitivity because an extended source with a $5\sigma$ peak flux density is more massive than a point source with the same peak flux density. These figures show that our data are consistent with previous studies.

Given the selection criteria discussed in the previous section, the completeness limit for a given noise level ($\sigma$) and source size ($r_{eff}$) can be estimated by measuring the fraction of the total flux density that is contained within an aperture set at 3/5 of the peak flux density, for a number of sources in the data. This fraction can then be applied to a source at the limit of the detection (i.e. a peak flux density equal to $5\sigma$), and gives the flux density that is contained within the $3\sigma$ contour. The source size ($r_{eff}$) is defined to be the radius that defines a circle with the same area as the elliptical aperture placed on the source. $\overline{r}_{eff}$ is calculated independently for each region and is taken to be the average size of the detected low-mass sources (based on the lowest mass one third of the sources). The low-mass sources are selected as being more representative of the average source size at the limit of detection.

\begin{figure*}
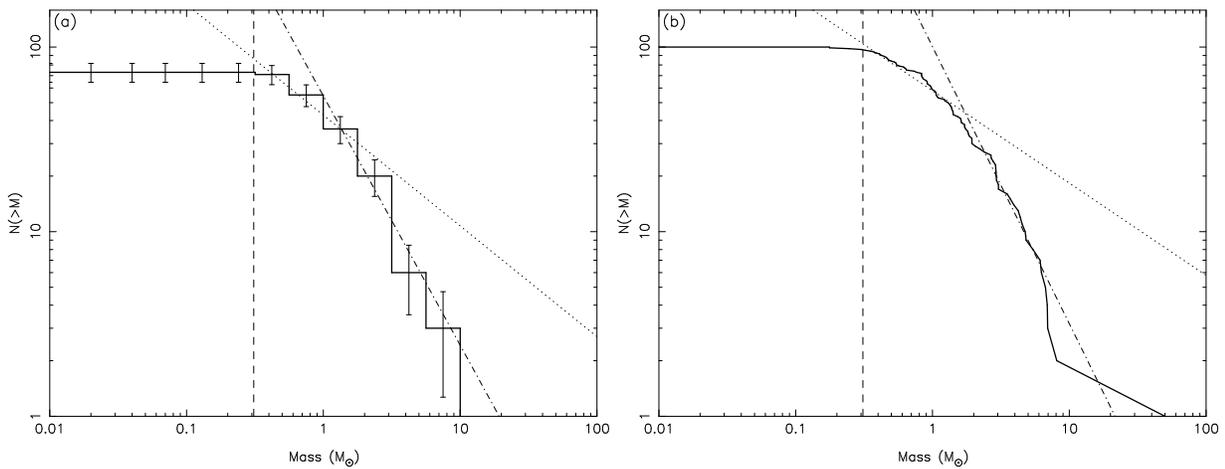

\begin{minipage}[t]{80mm}
\includegraphics[angle=270,width=80mm]{./fig5a.ps} 
\end{minipage}
\begin{minipage}[t]{80mm}
\includegraphics[angle=270,width=80mm]{./fig5b.ps}
\end{minipage}
\caption{a) The core mass function for the Orion BN region, plotted as a cumulative mass function to allow comparison with fig.~3 of \citet{2001A&A...372L..41M}. The Class I YSOs identified through the IRAC data have been removed from the data. The fits from \citeauthor{2001A&A...372L..41M} of $N(>m) \propto m^{-0.6}$ and $N(>m) \propto m^{-1.35}$ are shown as dotted and dot-dashed lines respectively. b) The same data, plotted as an unbinned cumulative mass function for comparison with fig.~8 of \citet{2001ApJ...559..307J}. All submillimetre detected sources are plotted. The fits from \citeauthor{2001ApJ...559..307J} of $N(>m) \propto m^{-0.5}$ and $N(>m) \propto m^{-1.5}$ are shown as dotted and dot-dashed lines respectively. The dotted line shows the position of the completeness limit.}
\label{OriBN_comparisons}
\end{figure*}

\begin{table}
\begin{center}
\caption{The point source sensitivity, completeness limit and average effective radius ($\overline{r}_{eff}$) for each of the four regions. $r_{eff}$ is based on the lowest mass 1/3 of the sources in that region.}
\label{table_stats}
\vspace{0.5cm}
\begin{tabular}{cccc} \hline
Cloud 		& $5\sigma$	& $\overline{r}_{eff}$	& Completeness 	\\
Name 		& sensitivity	& (arcsec)		& limit		\\ 
		& ($M_{\odot}$)	& 			&($M_{\odot}$)	\\ \hline
Orion AN	& 0.15		& 14			& 0.29		\\
Orion AS	& 0.13		& 16			& 0.37		\\ 
Orion BN	& 0.12		& 16			& 0.31		\\
Orion BS	& 0.17		& 17			& 0.45		\\ \hline
\end{tabular}
\end{center}
\end{table}

The relation between the completeness limit and the core mass function is insensitive to the assumptions of distance, temperature and dust opacity, discussed in the previous section, because both the core masses and the completeness limit scale in the same way if these assumptions are adjusted. The point source sensitivity, completeness limit and effective radius for each of the four regions are given in Table~\ref{table_stats}.

Looking at the Figures \ref{Orion_CMFs}(a-d), a number of features are apparent. The high mass side of the mass function declines as a power-law to higher masses, in agreement with previous studies of both Orion and other star-forming regions \citep{1998ApJ...508L..91T,1998A&A...336..150M,2001A&A...372L..41M,2001ApJ...559..307J,2002ApJ...575..950O,2006ApJ...639..259J}. The low-mass side of the mass function also declines to lower masses, and for two of the four regions mapped (Orion AN and Orion BN), this decline begins at significantly higher mass than the completeness limit. The peak of the core mass functions for these two regions show a peak in the mass bin that is centred on 1.3~$M_{\odot}$. This is a factor of 4 larger than the average 5$\sigma$ completeness limit of the two regions. 

We therefore believe that this is a real effect, and signifies the discovery of the turn-over in the core mass function. The physical significance of this is discussed in the following section. A third region (Orion AS) is also consistent with this picture, with the mass function appearing to turn over before the completeness limit is reached. However, there are a smaller number of sources detected in this region, and so there is more scatter in the different bins of the mass function. This causes the position of the peak of the mass function, and hence the significance of the turn-over, to be less certain. In the fourth region (Orion BS), there are a similarly small number of sources as in Orion AS. Also, the completeness limit is a factor of 1.5 times higher than in Orion AN and Orion BN. In this region there appears to be no significant decline in the number of cores until the completeness limit is reached.

As one would expect, there are no sources with a lower mass than the sensitivity limit, and in the mass functions for the Orion AN and Orion BN regions, there is a sharp decline in source numbers at the level of the completeness limit, lending support to our estimate of the completeness. 

Subsets of these data for the Orion BN region have been published previously \citep{2001ApJ...559..307J,2001A&A...372L..41M}. In order to allow easy comparison with these earlier works, mass functions for the Orion BN data are plotted in Figures~\ref{OriBN_comparisons}a \& b in the same cumulative form as that used in the earlier papers. The plotted fits to the data have the same slopes as are shown by \citeauthor{2001ApJ...559..307J} and \citeauthor{2001A&A...372L..41M} respectively. Though our data are complete to lower masses, Figures~\ref{OriBN_comparisons}a \& b show that the data presented here are consistent with the earlier work. 

\begin{figure}
\includegraphics[angle=270,width=83mm]{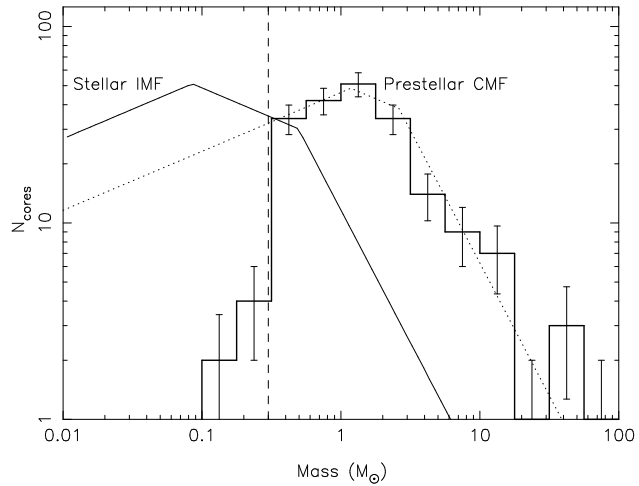}
\caption{The core mass function for the Orion AN and Orion BN regions. The average completeness limit for the two regions is shown as a dashed line. A three-part stellar IMF, normalised to the peak in N of the CMF, is overlaid as a thin solid line. The dotted line shows a three-part mass function with the same slopes as the IMF (see text for details).}
\label{comparison_imf}
\end{figure}

\section{Discussion}
In the previous section we demonstrated that the core mass functions for the two regions Orion AN and Orion BN turn over at a significantly higher mass than the completeness limit of the data. In Figure~\ref{comparison_imf}, we combine the data for these two regions and plot the resulting CMF on a log-log plot. 

The low-mass side of the CMF is best fitted by a power-law of following the form:
\begin{equation}
M\frac{dN}{dM} \propto M^{-x}
\label{imf1}
\end{equation}

where $x$ is equal to $-0.35 \pm 0.2$ between the completeness limit at ${\rm \sim 0.3 M_\odot}$ and the peak at ${\rm \sim 1.3~M_\odot}$. At higher masses than the peak, we find that the CMF is best fitted by a power-law with $x$ equal to $1.2 \pm 0.2$. 

We compare this CMF with the stellar IMF, for which we assume a three-part power-law form with the following exponents:
\begin{equation}
\begin{array}{lcrrll}
x & = & 1.35,  & 0.5~M_{\odot}  &<~M,~& \\
x & = & 0.3,   & 0.08~M_{\odot} &<~M~<& 0.5~M_{\odot}, \\
x & = & -0.3,  & ~~~~~~0.01~M_{\odot} &<~M~<& 0.08~M_{\odot}.
\end{array} 
\label{imf2}
\end{equation}
Above the peak of the IMF at $0.08~M_\odot$, the power-law slopes of the IMF are those observed for field stars, taken from \citet{2002Sci...295...82K}. Below $0.08~M_\odot$, the IMF is based on fits to the stellar populations of young clusters \citep{2003PASP..115..763C}. These clusters are observed to have power-law slopes varying between 0.2 and 0.4, therefore an average value of $0.3 \pm 0.1$ is used here. This IMF is plotted on Figure~\ref{comparison_imf} as a thin solid line.

The Orion CMF plotted in Figure~\ref{comparison_imf} is consistent with a mass function with the same slopes as the three-part power-law IMF given in Equations \ref{imf1} and \ref{imf2}, between the following masses.
\begin{equation}
\begin{array}{lcrrll}
x & = & 1.35,  & 2.4~M_{\odot}  &<~M,~& \\
x & = & 0.3,   & 1.3~M_{\odot} &<~M~<& 2.4~M_{\odot}, \\
x & = & -0.3,  & ~~~~~~0.4~M_{\odot} &<~M~<& 1.3~M_{\odot}.
\end{array} 
\label{imf3}
\end{equation}
This three-part power-law is plotted as a dotted line on Figure~\ref{comparison_imf}.

There are a number of similarities and differences between the CMF and the IMF shown in Figure~\ref{comparison_imf}. Both mass functions are well fit by negative power-law slopes at large masses, and decline to lower masses below a peak value. The most obvious difference between the two mass functions is that the turn-over of the CMF occurs at a much higher mass (${\rm \sim 1.3~M_\odot}$) than that of the IMF (${\rm \sim 0.08~M_\odot}$). 

If we were to assume a simple 1:1 mapping of the CMF to the IMF, whereby each core forms a single star, then these data show that the star formation efficiency (SFE) is only $\sim 6\%$ within the gravitationally bound core. The measured SFEs for star-forming regions are typically much larger, at $\sim 30\%$ \citep[][and references therein]{2003ARA&A..41...57L}. This discrepancy could be explained by the formation of multiple stars within each core.

The best fit power-law to the low-mass slope of the IMF has an exponent of $0.35 \pm 0.2$. This is consistent with the low-mass slope of the stellar IMF for young clusters which is observed to lie between 0.2 and 0.4 \citep{2003PASP..115..763C}. We compare the CMF to the IMFs of these systems because being so young, any causal relation is more likely to be apparent. The slope of the low-mass side of the CMF is based on only three data-points, and must therefore be considered a tentative result. However, if it is borne out to be correct by future observations, then it shows that the CMF continues to mimic the stellar IMF down to very low masses. We also note that a study of the nearer, lower-mass cloud $\rho$-Ophiuchi does not show a turn-over in the core mass function, down to their completeness limit of $\sim 0.1~M_\odot$ \citep{1998A&A...336..150M}. It is therefore possible that the position of the turn-over in the CMF varies as a function of environment.

\section{Conclusions}
We have combined all of the wide-field SCUBA data of the Orion molecular cloud from the JCMT archive, in order to create the deepest and most extensive maps of this region to date. We have extracted the cores from these data, and generated core mass functions for each region. We have estimated the completeness limit of each map, based on the sensitivity limits of the data and the average size of the low-mass cores. We find that for two of the regions, Orion A North and Orion B North, the CMF turns over at a significantly higher mass than the completeness limit. This marks the discovery of the turn-over of the CMF. We compare the CMF to a canonical stellar IMF, and find that the turn-over in the CMF occurs an a much higher mass than in the IMF. We find that the CMF is best fit by power-laws with exponents of 0.35 between 0.3 and 1.3 M$_\odot$, and -1.2 above 2.4 M$_\odot$. The CMF is consistent with the power-law slopes of the stellar IMF down to the completeness limit of the data.

\section*{Acknowledgements}
This research is based on observations obtained with the James Clerk Maxwell Telescope, which is operated by the Joint Astronomy Centre in Hilo, Hawaii on behalf of the parent organizations PPARC in the United Kingdom, the National Research Council of Canada and The Netherlands Organization for Scientific Research. The data were obtained using the Canadian Astronomy Data Centre, which is operated by the Herzberg Institute of Astrophysics, National Research Council of Canada. The observations were taken under the programs M98AC19, M98BC14, M98BH01, M98BN06, M98BU55, M99BC27, M99BC30, M00BC10, M00BC15, M01AC10 and M02AC32. The authors would like to thank the staff of the JCMT for assistance with the observations. This work is also based on observations made with the Spitzer Space Telescope, which is operated by the Jet Propulsion Laboratory, California Institute of Technology under a contract with NASA. DN acknowledges PPARC for PDRA support. This work was carried out while DWT was on sabbatical at the Observatoire de Bordeaux and at CEA Saclay, and he would like to thank both institutions for the hospitality accorded to him.

\appendix
\section{Tables}\label{appendix}
\begin{table*}
\begin{center}
\caption{The properties of each of the cores, determined from the submillimetre data. The source name is based on the region that it is found in and its J2000 coordinates. The semi-major and semi-minor dimensions are given for each source in columns 4 and 5. The peak and integrated flux densities for each source at 850 and 450~$\mu$m are given in columns 6 to 9. Dashes in columns 8 \& 9 indicate that the source is undetected at 5 sigma in the 450~$\mu$m data smoothed to a FWHM of 14 arcsec. The mass calculated from the 850~$\mu$m integrated flux density is given in column 10. Column 11 indicates the presence of a Class I protostar, detected using IRAC.}
\label{cores_table}
\begin{tabular}{|l|c|c|c|c|c|c|c|c|c|c|}\hline
&&& \multicolumn{2}{c|}{Aperture} & \multicolumn{2}{c|}{850~$\mu$m} & \multicolumn{2}{c|}{450~$\mu$m} & & \\ 
Source 	& RA 		& Dec.		& Semi-major& Semi-minor  & Peak	&Int.	& Peak		& Int. & Mass & YSO \\
Name	&(2000)		&(2000)		& (arcsec)  &(arcsec) &(Jy/beam)	&(Jy)	&(Jy/beam)	&(Jy)  & (M${_\odot}$) & \\ \hline
OrionAN-534497-54437 & 5:34:49.7 & -5:44:37 & 13 & 9 & 0.2 & 0.3 & -- & -- & 0.4 & N \\ 
OrionAN-534451-54421 & 5:34:45.1 & -5:44:21 & 18 & 17 & 0.2 & 0.6 & -- & -- & 0.9 & N \\ 
OrionAN-534594-54413 & 5:34:59.4 & -5:44:13 & 19 & 12 & 0.2 & 0.4 & 0.9 & 1.8 & 0.6 & N \\ 
OrionAN-534523-54412 & 5:34:52.3 & -5:44:12 & 21 & 20 & 0.3 & 0.9 & -- & -- & 1.4 & N \\ 
OrionAN-534548-54342 & 5:34:54.8 & -5:43:42 & 44 & 22 & 0.3 & 1.8 & 1.1 & 7.3 & 2.9 & N \\ 
OrionAN-534575-54338 & 5:34:57.5 & -5:43:38 & 16 & 9 & 0.2 & 0.3 & 0.9 & 1.5 & 0.5 & N \\ 
OrionAN-534576-54157 & 5:34:57.6 & -5:41:57 & 41 & 22 & 0.3 & 1.9 & 1.1 & 6.4 & 2.9 & Y \\ 
OrionAN-534439-54128 & 5:34:43.9 & -5:41:28 & 21 & 20 & 0.4 & 0.8 & 0.8 & 1.1 & 1.3 & Y \\ 
OrionAN-534581-54056 & 5:34:58.1 & -5:40:56 & 23 & 21 & 0.3 & 1.0 & 1.1 & 3.8 & 1.6 & N \\ 
OrionAN-535001-54012 & 5:35:00.1 & -5:40:12 & 29 & 16 & 0.3 & 0.9 & 1.0 & 2.8 & 1.4 & N \\ 
OrionAN-534507-53856 & 5:34:50.7 & -5:38:56 & 30 & 17 & 0.2 & 0.7 & 0.7 & 1.5 & 1.1 & N \\ 
OrionAN-535001-53857 & 5:35:00.1 & -5:38:57 & 32 & 22 & 0.4 & 2.1 & 1.5 & 7.0 & 3.2 & N \\ 
OrionAN-535021-53805 & 5:35:02.1 & -5:38:05 & 31 & 22 & 0.5 & 2.3 & 2.0 & 8.7 & 3.7 & Y \\ 
OrionAN-535100-53800 & 5:35:10.0 & -5:38:00 & 32 & 21 & 0.3 & 1.3 & 1.1 & 4.3 & 2.0 & N \\ 
OrionAN-534510-53757 & 5:34:51.0 & -5:37:57 & 20 & 13 & 0.1 & 0.2 & -- & -- & 0.3 & N \\ 
OrionAN-535049-53724 & 5:35:04.9 & -5:37:24 & 30 & 18 & 0.8 & 2.2 & 2.9 & 8.2 & 3.4 & Y \\ 
OrionAN-535023-53714 & 5:35:02.3 & -5:37:14 & 18 & 12 & 0.4 & 0.8 & 1.5 & 3.4 & 1.3 & Y \\ 
OrionAN-534538-53715 & 5:34:53.8 & -5:37:15 & 24 & 14 & 0.2 & 0.4 & 0.5 & 1.0 & 0.6 & N \\ 
OrionAN-535016-53648 & 5:35:01.6 & -5:36:48 & 18 & 14 & 0.4 & 0.9 & 1.6 & 4.0 & 1.4 & N \\ 
OrionAN-535296-53644 & 5:35:29.6 & -5:36:44 & 52 & 29 & 0.2 & 1.6 & 0.8 & 5.3 & 2.5 & N \\ 
OrionAN-534569-53640 & 5:34:56.9 & -5:36:40 & 35 & 20 & 0.2 & 1.0 & 0.9 & 4.1 & 1.6 & N \\ 
OrionAN-535028-53613 & 5:35:02.8 & -5:36:13 & 32 & 24 & 0.7 & 3.0 & 3.0 & 12.8 & 4.8 & N \\ 
OrionAN-535082-53558 & 5:35:08.2 & -5:35:58 & 34 & 25 & 1.0 & 3.3 & 3.7 & 12.4 & 5.2 & Y \\ 
OrionAN-535103-53507 & 5:35:10.3 & -5:35:07 & 25 & 22 & 0.7 & 2.2 & 2.3 & 7.7 & 3.4 & Y \\ 
OrionAN-535046-53459 & 5:35:04.6 & -5:34:59 & 41 & 22 & 0.7 & 3.4 & 2.9 & 12.9 & 5.3 & Y \\ 
OrionAN-535119-53430 & 5:35:11.9 & -5:34:30 & 25 & 21 & 0.7 & 2.0 & 2.5 & 7.5 & 3.1 & N \\ 
OrionAN-535064-53400 & 5:35:06.4 & -5:34:00 & 17 & 13 & 0.4 & 0.7 & 1.3 & 2.8 & 1.2 & Y \\ 
OrionAN-534562-53401 & 5:34:56.2 & -5:34:01 & 17 & 14 & 0.2 & 0.3 & 0.6 & 1.1 & 0.5 & N \\ 
OrionAN-535058-53310 & 5:35:05.8 & -5:33:10 & 29 & 19 & 0.5 & 1.5 & 1.8 & 5.4 & 2.4 & N \\ 
OrionAN-535149-53307 & 5:35:14.9 & -5:33:07 & 17 & 13 & 0.2 & 0.3 & 0.6 & 1.1 & 0.5 & Y \\ 
OrionAN-535178-53020 & 5:35:17.8 & -5:30:20 & 67 & 33 & 0.3 & 3.6 & 2.1 & 26.5 & 5.7 & N \\ 
OrionAN-535129-53003 & 5:35:12.9 & -5:30:03 & 16 & 13 & 0.2 & 0.3 & 0.7 & 1.0 & 0.5 & N \\ 
OrionAN-535093-52958 & 5:35:09.3 & -5:29:58 & 15 & 9 & 0.2 & 0.2 & 0.6 & 0.7 & 0.3 & N \\ 
OrionAN-535106-52926 & 5:35:10.6 & -5:29:26 & 20 & 19 & 0.2 & 0.3 & 0.7 & 1.9 & 0.5 & N \\ 
OrionAN-535185-52830 & 5:35:18.5 & -5:28:30 & 26 & 18 & 0.6 & 1.5 & 1.7 & 4.9 & 2.3 & Y \\ 
OrionAN-535274-52757 & 5:35:27.4 & -5:27:57 & 27 & 20 & 0.3 & 1.3 & -- & -- & 2.1 & N \\ 
OrionAN-535025-52754 & 5:35:02.5 & -5:27:54 & 27 & 9 & 0.3 & 0.6 & 0.7 & 1.4 & 0.9 & N \\ 
OrionAN-535146-52623 & 5:35:14.6 & -5:26:23 & 71 & 36 & 2.4 & 34.3 & 9.6 & 89.4 & 53.9 & Y \\ 
OrionAN-535270-52523 & 5:35:27.0 & -5:25:23 & 18 & 17 & 0.5 & 1.1 & 1.3 & 3.4 & 1.7 & N \\ 
OrionAN-535222-52510 & 5:35:22.2 & -5:25:10 & 35 & 26 & 2.5 & 16.1 & 7.8 & 48.7 & 44.0 & N \\ 
OrionAN-535022-52453 & 5:35:02.2 & -5:24:53 & 29 & 14 & 0.4 & 0.9 & -- & -- & 1.3 & N \\ 
OrionAN-535254-52433 & 5:35:25.4 & -5:24:33 & 29 & 22 & 3.8 & 13.9 & 11.1 & 41.2 & 50.7 & N \\ 
OrionAN-535049-52420 & 5:35:04.9 & -5:24:20 & 25 & 14 & 0.6 & 1.1 & 1.9 & 2.2 & 1.8 & N \\ 
OrionAN-535136-52409 & 5:35:13.6 & -5:24:09 & 47 & 38 & 46.3 & 213.5 & 137.5 & 629.0 & 392.9 & Y \\ 
OrionAN-535295-52400 & 5:35:29.5 & -5:24:00 & 16 & 10 & 0.4 & 0.5 & 1.4 & 1.7 & 0.9 & N \\ 
OrionAN-535030-52328 & 5:35:03.0 & -5:23:28 & 13 & 11 & 0.3 & 0.5 & -- & -- & 0.8 & N \\ 
OrionAN-535220-52305 & 5:35:22.0 & -5:23:05 & 18 & 10 & 0.2 & 0.4 & -- & -- & 0.6 & N \\ 
OrionAN-535056-52253 & 5:35:05.6 & -5:22:53 & 40 & 18 & 1.6 & 5.0 & 7.8 & 30.5 & 7.9 & N \\ 
OrionAN-535319-52233 & 5:35:31.9 & -5:22:33 & 39 & 22 & 1.2 & 3.1 & 3.2 & 4.8 & 4.9 & N \\ 
OrionAN-535023-52231 & 5:35:02.3 & -5:22:31 & 10 & 9 & 0.7 & 0.9 & 2.5 & 3.4 & 1.4 & N \\ 
OrionAN-535264-52230 & 5:35:26.4 & -5:22:30 & 37 & 18 & 1.3 & 4.8 & 3.3 & 11.0 & 7.5 & N \\ 
OrionAN-535144-52233 & 5:35:14.4 & -5:22:33 & 59 & 48 & 143.5 & 517.8 & 279.2 & 1263.8 & 991.4 & N \\ 
OrionAN-535241-52223 & 5:35:24.1 & -5:22:23 & 25 & 16 & 1.1 & 3.2 & 3.2 & 8.2 & 5.1 & Y \\ 
OrionAN-535031-52140 & 5:35:03.1 & -5:21:40 & 23 & 10 & 0.3 & 0.4 & 1.3 & 2.2 & 0.6 & N \\ 
OrionAN-535000-52145 & 5:35:00.0 & -5:21:45 & 35 & 20 & 1.0 & 2.7 & 4.3 & 11.0 & 4.3 & N \\ 
OrionAN-535269-52129 & 5:35:26.9 & -5:21:29 & 18 & 9 & 0.6 & 1.1 & -- & -- & 1.7 & N \\ \hline
\end{tabular}
\end{center}
\end{table*}
\begin{table*}
\begin{center}
\contcaption{Core properties}
\begin{tabular}{|l|c|c|c|c|c|c|c|c|c|c|}\hline
&&& \multicolumn{2}{c|}{Aperture} & \multicolumn{2}{c|}{850~$\mu$m} & \multicolumn{2}{c|}{450~$\mu$m} & & \\ 
Source 	& RA 		& Dec.		& Semi-major& Semi-minor  & Peak	&Int.	& Peak		& Int. & Mass & YSO \\
Name	&(2000)		&(2000)		& (arcsec)  &(arcsec) &(Jy/beam)	&(Jy)	&(Jy/beam)	&(Jy)  & (M${_\odot}$) & \\ \hline
OrionAN-535354-52130 & 5:35:35.4 & -5:21:30 & 10 & 8 & 0.2 & 0.1 & -- & -- & 0.1 & N \\ 
OrionAN-535327-52126 & 5:35:32.7 & -5:21:26 & 17 & 10 & 0.4 & 0.7 & -- & -- & 1.1 & N \\ 
OrionAN-535181-52129 & 5:35:18.1 & -5:21:29 & 26 & 14 & 5.4 & 12.7 & 17.3 & 39.8 & 36.4 & N \\ 
OrionAN-535059-52121 & 5:35:05.9 & -5:21:21 & 28 & 18 & 0.8 & 2.1 & 4.2 & 13.5 & 3.3 & N \\ 
OrionAN-535161-52111 & 5:35:16.1 & -5:21:11 & 19 & 16 & 6.0 & 15.6 & 23.7 & 63.2 & 42.6 & Y \\ 
OrionAN-535425-52057 & 5:35:42.5 & -5:20:57 & 30 & 16 & 0.5 & 1.3 & 1.8 & 4.2 & 2.0 & N \\ 
OrionAN-535184-52049 & 5:35:18.4 & -5:20:49 & 23 & 15 & 4.0 & 9.1 & 13.8 & 31.8 & 26.4 & Y \\ 
OrionAN-534576-52053 & 5:34:57.6 & -5:20:53 & 27 & 15 & 0.4 & 0.9 & 1.3 & 3.4 & 1.5 & N \\ 
OrionAN-535314-52048 & 5:35:31.4 & -5:20:48 & 45 & 22 & 1.2 & 8.2 & 4.2 & 29.5 & 12.8 & N \\ 
OrionAN-535040-52046 & 5:35:04.0 & -5:20:46 & 21 & 11 & 0.3 & 0.7 & 1.3 & 2.9 & 1.1 & N \\ 
OrionAN-535163-52043 & 5:35:16.3 & -5:20:43 & 23 & 22 & 7.6 & 24.8 & 30.8 & 93.2 & 62.1 & N \\ 
OrionAN-535091-52027 & 5:35:09.1 & -5:20:27 & 45 & 25 & 1.4 & 4.4 & 7.3 & 28.6 & 6.9 & N \\ 
OrionAN-535054-52025 & 5:35:05.4 & -5:20:25 & 14 & 10 & 0.3 & 0.4 & 1.4 & 2.0 & 0.7 & Y \\ 
OrionAN-535413-52021 & 5:35:41.3 & -5:20:21 & 16 & 16 & 0.4 & 0.6 & 1.2 & 2.4 & 1.0 & Y \\ 
OrionAN-535256-52015 & 5:35:25.6 & -5:20:15 & 40 & 24 & 1.5 & 8.3 & 5.9 & 31.9 & 13.0 & N \\ 
OrionAN-535010-51936 & 5:35:01.0 & -5:19:36 & 25 & 15 & 0.2 & 0.6 & 0.7 & 1.1 & 0.9 & N \\ 
OrionAN-535398-51925 & 5:35:39.8 & -5:19:25 & 11 & 9 & 0.3 & 0.3 & -- & -- & 0.4 & N \\ 
OrionAN-535333-51920 & 5:35:33.3 & -5:19:20 & 18 & 7 & 0.5 & 0.8 & -- & -- & 1.2 & N \\ 
OrionAN-535253-51917 & 5:35:25.3 & -5:19:17 & 13 & 8 & 0.5 & 0.7 & -- & -- & 1.1 & N \\ 
OrionAN-535170-51924 & 5:35:17.0 & -5:19:24 & 49 & 29 & 7.5 & 40.6 & 30.8 & 163.4 & 86.7 & Y \\ 
OrionAN-535399-51905 & 5:35:39.9 & -5:19:05 & 11 & 7 & 0.3 & 0.2 & -- & -- & 0.3 & N \\ 
OrionAN-535097-51849 & 5:35:09.7 & -5:18:49 & 22 & 15 & 0.5 & 1.1 & 2.6 & 6.7 & 1.7 & N \\ 
OrionAN-535146-51847 & 5:35:14.6 & -5:18:47 & 27 & 14 & 3.6 & 8.5 & 15.5 & 38.7 & 24.4 & N \\ 
OrionAN-535358-51838 & 5:35:35.8 & -5:18:38 & 9 & 8 & 0.5 & 0.7 & -- & -- & 1.0 & N \\ 
OrionAN-535225-51834 & 5:35:22.5 & -5:18:34 & 15 & 9 & 0.6 & 0.9 & -- & -- & 1.5 & N \\ 
OrionAN-535112-51828 & 5:35:11.2 & -5:18:28 & 14 & 13 & 0.4 & 0.6 & 2.0 & 3.9 & 1.0 & N \\ 
OrionAN-535191-51829 & 5:35:19.1 & -5:18:29 & 12 & 8 & 2.1 & 2.9 & 8.2 & 11.5 & 11.6 & N \\ 
OrionAN-535376-51822 & 5:35:37.6 & -5:18:22 & 29 & 18 & 0.8 & 3.4 & 3.2 & 13.8 & 5.4 & N \\ 
OrionAN-535011-51817 & 5:35:01.1 & -5:18:17 & 16 & 11 & 0.2 & 0.4 & 0.7 & 1.2 & 0.6 & N \\ 
OrionAN-535200-51810 & 5:35:20.0 & -5:18:10 & 20 & 14 & 2.0 & 4.7 & 7.5 & 15.3 & 14.2 & N \\ 
OrionAN-535036-51801 & 5:35:03.6 & -5:18:01 & 19 & 12 & 0.4 & 0.5 & 1.2 & 2.0 & 0.9 & N \\ 
OrionAN-535091-51752 & 5:35:09.1 & -5:17:52 & 32 & 15 & 0.5 & 1.5 & 2.4 & 7.9 & 2.3 & N \\ 
OrionAN-535385-51718 & 5:35:38.5 & -5:17:18 & 28 & 18 & 0.5 & 1.8 & 2.0 & 7.2 & 2.8 & N \\ 
OrionAN-535068-51709 & 5:35:06.8 & -5:17:09 & 15 & 9 & 0.3 & 0.4 & 1.2 & 1.9 & 0.6 & N \\ 
OrionAN-535104-51651 & 5:35:10.4 & -5:16:51 & 18 & 16 & 0.3 & 0.6 & 1.5 & 3.5 & 1.0 & N \\ 
OrionAN-535149-51646 & 5:35:14.9 & -5:16:46 & 14 & 11 & 0.3 & 0.4 & 1.3 & 2.1 & 0.6 & Y \\ 
OrionAN-534570-51620 & 5:34:57.0 & -5:16:20 & 21 & 13 & 0.3 & 0.7 & 1.1 & 2.7 & 1.0 & N \\ 
OrionAN-535006-51620 & 5:35:00.6 & -5:16:20 & 32 & 16 & 0.4 & 1.5 & 1.9 & 6.9 & 2.4 & N \\ 
OrionAN-535150-51613 & 5:35:15.0 & -5:16:13 & 22 & 17 & 0.4 & 1.1 & 1.7 & 4.6 & 1.7 & N \\ 
OrionAN-535366-51559 & 5:35:36.6 & -5:15:59 & 51 & 26 & 0.7 & 4.4 & 2.3 & 17.9 & 6.9 & N \\ 
OrionAN-535063-51550 & 5:35:06.3 & -5:15:50 & 16 & 13 & 0.2 & 0.4 & 0.9 & 2.0 & 0.7 & N \\ 
OrionAN-535246-51538 & 5:35:24.6 & -5:15:38 & 20 & 17 & 0.4 & 1.0 & 1.2 & 3.6 & 1.6 & N \\ 
OrionAN-535196-51535 & 5:35:19.6 & -5:15:35 & 27 & 17 & 1.3 & 2.0 & 4.4 & 6.5 & 3.2 & Y \\ 
OrionAN-535020-51525 & 5:35:02.0 & -5:15:25 & 37 & 18 & 0.5 & 1.5 & 2.1 & 7.2 & 2.4 & N \\ 
OrionAN-535101-51518 & 5:35:10.1 & -5:15:18 & 14 & 11 & 0.2 & 0.2 & 0.7 & 1.0 & 0.4 & N \\ 
OrionAN-535159-51509 & 5:35:15.9 & -5:15:09 & 38 & 20 & 0.4 & 1.8 & 1.6 & 6.2 & 2.8 & Y \\ 
OrionAN-535387-51508 & 5:35:38.7 & -5:15:08 & 20 & 15 & 0.2 & 0.6 & 0.6 & 1.2 & 0.9 & N \\ 
OrionAN-535214-51458 & 5:35:21.4 & -5:14:58 & 24 & 14 & 0.8 & 1.9 & 3.3 & 7.7 & 3.0 & N \\ 
OrionAN-535218-51422 & 5:35:21.8 & -5:14:22 & 20 & 19 & 0.8 & 2.0 & 3.1 & 7.9 & 3.2 & Y \\ 
OrionAN-535189-51412 & 5:35:18.9 & -5:14:12 & 16 & 10 & 0.3 & 0.5 & 0.7 & 0.7 & 0.7 & N \\ 
OrionAN-535350-51357 & 5:35:35.0 & -5:13:57 & 26 & 13 & 0.2 & 0.4 & 0.8 & 1.5 & 0.6 & N \\ 
OrionAN-535183-51338 & 5:35:18.3 & -5:13:38 & 13 & 8 & 0.3 & 0.3 & -- & -- & 0.5 & N \\ 
OrionAN-535014-51340 & 5:35:01.4 & -5:13:40 & 48 & 22 & 0.3 & 0.9 & 1.0 & 3.6 & 1.4 & N \\ 
OrionAN-535157-51318 & 5:35:15.7 & -5:13:18 & 12 & 9 & 0.2 & 0.3 & 0.5 & 0.6 & 0.4 & N \\ 
OrionAN-535217-51312 & 5:35:21.7 & -5:13:12 & 18 & 18 & 1.8 & 4.1 & 5.4 & 12.9 & 10.1 & N \\ 
OrionAN-535229-51240 & 5:35:22.9 & -5:12:40 & 30 & 18 & 2.1 & 8.3 & 7.0 & 28.3 & 17.3 & N \\ 
OrionAN-535290-51209 & 5:35:29.0 & -5:12:09 & 25 & 14 & 0.3 & 0.7 & 1.2 & 2.1 & 1.1 & N \\ 
OrionAN-535183-51207 & 5:35:18.3 & -5:12:07 & 13 & 9 & 0.2 & 0.2 & 0.6 & 0.5 & 0.3 & N \\ 
OrionAN-535161-51207 & 5:35:16.1 & -5:12:07 & 30 & 21 & 0.3 & 1.0 & 1.0 & 4.1 & 1.6 & N \\ 
OrionAN-535234-51205 & 5:35:23.4 & -5:12:05 & 21 & 16 & 1.4 & 3.2 & 4.9 & 11.5 & 8.0 & Y \\ \hline
\end{tabular}
\end{center}
\end{table*}
\begin{table*}
\begin{center}
\contcaption{Core properties}
\begin{tabular}{|l|c|c|c|c|c|c|c|c|c|c|}\hline
&&& \multicolumn{2}{c|}{Aperture} & \multicolumn{2}{c|}{850~$\mu$m} & \multicolumn{2}{c|}{450~$\mu$m} & & \\ 
Source 	& RA 		& Dec.		& Semi-major& Semi-minor  & Peak	&Int.	& Peak		& Int. & Mass & YSO \\
Name	&(2000)		&(2000)		& (arcsec)  &(arcsec) &(Jy/beam)	&(Jy)	&(Jy/beam)	&(Jy)  & (M${_\odot}$) & \\ \hline
OrionAN-535271-51139 & 5:35:27.1 & -5:11:39 & 23 & 14 & 0.6 & 1.3 & 1.9 & 4.1 & 2.0 & N \\ 
OrionAN-535154-51114 & 5:35:15.4 & -5:11:14 & 20 & 12 & 0.2 & 0.3 & -- & -- & 0.4 & N \\ 
OrionAN-535216-51039 & 5:35:21.6 & -5:10:39 & 19 & 12 & 0.3 & 0.6 & 1.3 & 2.2 & 0.9 & N \\ 
OrionAN-535279-51025 & 5:35:27.9 & -5:10:25 & 11 & 10 & 1.2 & 1.9 & -- & -- & 4.2 & N \\ 
OrionAN-535138-51013 & 5:35:13.8 & -5:10:13 & 31 & 19 & 0.2 & 0.7 & 0.6 & 2.0 & 1.1 & N \\ 
OrionAN-535225-51014 & 5:35:22.5 & -5:10:14 & 17 & 13 & 0.5 & 0.9 & 2.1 & 3.9 & 1.4 & N \\ 
OrionAN-535265-51011 & 5:35:26.5 & -5:10:11 & 31 & 22 & 6.8 & 17.0 & 23.2 & 62.1 & 33.3 & Y \\ 
OrionAN-535243-50937 & 5:35:24.3 & -5:09:37 & 16 & 11 & 0.4 & 0.7 & 1.4 & 2.5 & 1.5 & N \\ 
OrionAN-535128-50936 & 5:35:12.8 & -5:09:36 & 15 & 10 & 0.2 & 0.2 & -- & -- & 0.4 & N \\ 
OrionAN-535276-50935 & 5:35:27.6 & -5:09:35 & 23 & 17 & 3.2 & 9.0 & 11.2 & 32.2 & 17.1 & Y \\ 
OrionAN-535185-50935 & 5:35:18.5 & -5:09:35 & 42 & 10 & 0.2 & 0.1 & 0.9 & 1.4 & 0.2 & N \\ 
OrionAN-535305-50926 & 5:35:30.5 & -5:09:26 & 11 & 9 & 0.2 & 0.2 & 0.5 & 0.6 & 0.3 & N \\ 
OrionAN-535264-50830 & 5:35:26.4 & -5:08:30 & 16 & 14 & 0.8 & 1.8 & 3.4 & 8.3 & 2.8 & N \\ 
OrionAN-535245-50832 & 5:35:24.5 & -5:08:32 & 26 & 16 & 1.7 & 4.3 & 5.0 & 15.0 & 6.8 & N \\ 
OrionAN-535350-50821 & 5:35:35.0 & -5:08:21 & 11 & 10 & 0.2 & 0.2 & 0.5 & 0.7 & 0.4 & N \\ 
OrionAN-535222-50817 & 5:35:22.2 & -5:08:17 & 18 & 14 & 0.5 & 1.2 & 2.2 & 5.4 & 1.9 & Y \\ 
OrionAN-535276-50802 & 5:35:27.6 & -5:08:02 & 21 & 12 & 0.4 & 0.7 & 1.5 & 3.6 & 1.2 & N \\ 
OrionAN-535245-50754 & 5:35:24.5 & -5:07:54 & 21 & 13 & 1.8 & 4.0 & 5.3 & 13.3 & 6.2 & Y \\ 
OrionAN-535235-50734 & 5:35:23.5 & -5:07:34 & 15 & 13 & 1.4 & 2.8 & 5.2 & 10.7 & 4.4 & N \\ 
OrionAN-535263-50729 & 5:35:26.3 & -5:07:29 & 14 & 10 & 0.3 & 0.4 & 1.1 & 1.8 & 0.7 & N \\ 
OrionAN-535279-50711 & 5:35:27.9 & -5:07:11 & 18 & 14 & 0.5 & 0.8 & 1.6 & 2.8 & 1.3 & N \\ 
OrionAN-535236-50711 & 5:35:23.6 & -5:07:11 & 16 & 13 & 1.2 & 2.5 & 4.4 & 9.3 & 4.0 & N \\ 
OrionAN-535403-50648 & 5:35:40.3 & -5:06:48 & 16 & 14 & 0.3 & 0.5 & 0.9 & 1.9 & 0.8 & Y \\ 
OrionAN-535324-50633 & 5:35:32.4 & -5:06:33 & 19 & 11 & 0.4 & 0.7 & 1.4 & 2.8 & 1.1 & N \\ 
OrionAN-535394-50627 & 5:35:39.4 & -5:06:27 & 24 & 13 & 0.3 & 0.7 & 1.0 & 2.7 & 1.1 & N \\ 
OrionAN-535376-50616 & 5:35:37.6 & -5:06:16 & 27 & 14 & 0.2 & 0.7 & 0.9 & 2.7 & 1.1 & N \\ 
OrionAN-535171-50603 & 5:35:17.1 & -5:06:03 & 23 & 20 & 0.4 & 1.3 & 1.0 & 3.7 & 2.0 & Y \\ 
OrionAN-535258-50551 & 5:35:25.8 & -5:05:51 & 38 & 22 & 1.9 & 6.3 & 6.1 & 19.4 & 9.9 & Y \\ 
OrionAN-535324-50547 & 5:35:32.4 & -5:05:47 & 50 & 28 & 0.9 & 5.3 & 2.7 & 15.5 & 8.4 & Y \\ 
OrionAN-535274-50511 & 5:35:27.4 & -5:05:11 & 32 & 23 & 1.8 & 5.3 & 6.0 & 17.1 & 8.3 & Y \\ 
OrionAN-535192-50457 & 5:35:19.2 & -5:04:57 & 51 & 27 & 0.8 & 5.9 & 2.9 & 15.4 & 9.3 & Y \\ 
OrionAN-535301-50453 & 5:35:30.1 & -5:04:53 & 12 & 9 & 0.3 & 0.4 & 1.1 & 1.4 & 0.6 & N \\ 
OrionAN-535265-50356 & 5:35:26.5 & -5:03:56 & 37 & 28 & 1.9 & 6.5 & 6.2 & 26.0 & 10.2 & Y \\ 
OrionAN-535255-50237 & 5:35:25.5 & -5:02:37 & 27 & 15 & 0.9 & 2.2 & 3.3 & 8.4 & 3.5 & N \\ 
OrionAN-535175-50234 & 5:35:17.5 & -5:02:34 & 26 & 19 & 0.4 & 1.1 & 1.3 & 4.5 & 1.8 & N \\ 
OrionAN-535414-50231 & 5:35:41.4 & -5:02:31 & 19 & 18 & 0.4 & 1.0 & 1.3 & 3.2 & 1.6 & N \\ 
OrionAN-535203-50147 & 5:35:20.3 & -5:01:47 & 18 & 16 & 0.7 & 1.4 & 2.2 & 5.2 & 2.2 & N \\ 
OrionAN-535182-50147 & 5:35:18.2 & -5:01:47 & 21 & 18 & 0.7 & 1.8 & 2.5 & 7.1 & 2.8 & N \\ 
OrionAN-535235-50132 & 5:35:23.5 & -5:01:32 & 21 & 16 & 6.5 & 9.7 & 15.3 & 26.6 & 29.8 & Y \\ 
OrionAN-535261-50126 & 5:35:26.1 & -5:01:26 & 26 & 12 & 0.5 & 1.4 & 1.8 & 5.6 & 2.2 & N \\ 
OrionAN-535224-50114 & 5:35:22.4 & -5:01:14 & 13 & 10 & 2.7 & 3.9 & 8.6 & 12.8 & 12.3 & Y \\ 
OrionAN-535383-50100 & 5:35:38.3 & -5:01:00 & 26 & 12 & 0.2 & 0.3 & 0.6 & 1.3 & 0.5 & N \\ 
OrionAN-535207-50053 & 5:35:20.7 & -5:00:53 & 17 & 13 & 2.7 & 5.5 & 9.0 & 19.1 & 14.7 & N \\ 
OrionAN-535316-50040 & 5:35:31.6 & -5:00:40 & 14 & 12 & 0.3 & 0.4 & 0.9 & 1.1 & 0.6 & N \\ 
OrionAN-535182-50021 & 5:35:18.2 & -5:00:21 & 19 & 14 & 2.9 & 5.9 & 8.9 & 19.1 & 15.7 & Y \\ 
OrionAN-535112-50011 & 5:35:11.2 & -5:00:11 & 26 & 18 & 0.5 & 1.7 & 2.0 & 7.0 & 2.7 & N \\ 
OrionAN-535161-50000 & 5:35:16.1 & -5:00:00 & 16 & 13 & 1.4 & 2.9 & 5.3 & 11.2 & 7.7 & Y \\ 
OrionAN-535344-45952 & 5:35:34.4 & -4:59:52 & 21 & 14 & 0.3 & 0.6 & 1.1 & 2.5 & 0.9 & Y \\ 
OrionAN-535300-45946 & 5:35:30.0 & -4:59:46 & 19 & 16 & 0.6 & 1.1 & 1.8 & 4.1 & 1.7 & Y \\ 
OrionAN-535034-45908 & 5:35:03.4 & -4:59:08 & 14 & 8 & 0.2 & 0.3 & -- & -- & 0.5 & N \\ 
OrionAN-535296-45836 & 5:35:29.6 & -4:58:36 & 48 & 22 & 1.5 & 4.7 & 4.7 & 17.1 & 7.4 & Y \\ 
OrionAN-535005-45831 & 5:35:00.5 & -4:58:31 & 24 & 9 & 0.3 & 0.6 & 1.0 & 2.0 & 0.9 & N \\ 
OrionAN-535046-45724 & 5:35:04.6 & -4:57:24 & 32 & 23 & 0.4 & 1.9 & 1.4 & 7.2 & 3.0 & N \\ 
OrionAN-535226-45633 & 5:35:22.6 & -4:56:33 & 17 & 15 & 0.2 & 0.3 & -- & -- & 0.5 & N \\ 
OrionAN-535177-45634 & 5:35:17.7 & -4:56:34 & 18 & 15 & 0.3 & 0.7 & 1.1 & 2.3 & 1.0 & N \\ 
OrionAN-535067-45631 & 5:35:06.7 & -4:56:31 & 28 & 23 & 0.8 & 2.3 & 2.2 & 7.3 & 3.5 & N \\ 
OrionAS-536186-64737 & 5:36:18.6 & -6:47:37 & 24 & 18 & 0.1 & 0.5 & 0.7 & 3.1 & 0.8 & N \\ 
OrionAS-536273-64730 & 5:36:27.3 & -6:47:30 & 36 & 24 & 0.2 & 1.6 & 1.0 & 7.9 & 2.5 & N \\ 
OrionAS-536229-64618 & 5:36:22.9 & -6:46:18 & 52 & 30 & 1.7 & 6.2 & 5.7 & 23.4 & 9.8 & Y \\ 
OrionAS-536179-64544 & 5:36:17.9 & -6:45:44 & 38 & 30 & 1.1 & 4.5 & 3.7 & 17.0 & 7.0 & N \\ \hline
\end{tabular}
\end{center}
\end{table*}
\begin{table*}
\begin{center}
\contcaption{Core properties}
\begin{tabular}{|l|c|c|c|c|c|c|c|c|c|c|}\hline
&&& \multicolumn{2}{c|}{Aperture} & \multicolumn{2}{c|}{850~$\mu$m} & \multicolumn{2}{c|}{450~$\mu$m} & & \\ 
Source 	& RA 		& Dec.		& Semi-major& Semi-minor  & Peak	&Int.	& Peak		& Int. & Mass & YSO \\
Name	&(2000)		&(2000)		& (arcsec)  &(arcsec) &(Jy/beam)	&(Jy)	&(Jy/beam)	&(Jy)  & (M${_\odot}$) & \\ \hline
OrionAS-536259-64440 & 5:36:25.9 & -6:44:40 & 38 & 27 & 0.5 & 2.2 & 1.3 & 7.1 & 3.5 & N \\ 
OrionAS-536358-64008 & 5:36:35.8 & -6:40:08 & 23 & 12 & 0.1 & 0.34 & -- & -- & 0.5 & N \\ 
OrionAS-536361-63857 & 5:36:36.1 & -6:38:57 & 38 & 18 & 1.0 & 1.8 & 2.0 & 4.2 & 2.8 & Y \\ 
OrionAS-536332-63812 & 5:36:33.2 & -6:38:12 & 19 & 14 & 0.2 & 0.4 & 0.6 & 1.3 & 0.7 & N \\ 
OrionAS-536170-63757 & 5:36:17.0 & -6:37:57 & 20 & 17 & 0.3 & 0.7 & 0.4 & 1.3 & 1.1 & N \\ 
OrionAS-537004-63715 & 5:37:00.4 & -6:37:15 & 36 & 19 & 0.5 & 1.5 & 1.1 & 4.6 & 2.4 & Y \\ 
OrionAS-535516-63456 & 5:35:51.6 & -6:34:56 & 24 & 16 & 0.5 & 1.2 & 0.7 & 2.5 & 1.9 & N \\ 
OrionAS-536119-63350 & 5:36:11.9 & -6:33:50 & 21 & 15 & 0.1 & 0.5 & 0.4 & 0.7 & 0.8 & N \\ 
OrionAS-536581-63114 & 5:36:58.1 & -6:31:14 & 12 & 11 & 0.2 & 0.3 & 1.0 & 1.5 & 0.5 & N \\ 
OrionAS-536518-62946 & 5:36:51.8 & -6:29:46 & 31 & 24 & 0.2 & 0.8 & 1.1 & 5.0 & 1.3 & Y \\ 
OrionAS-536193-62905 & 5:36:19.3 & -6:29:05 & 16 & 14 & 0.1 & 0.2 & 0.5 & 1.2 & 0.3 & N \\ 
OrionAS-536404-62848 & 5:36:40.4 & -6:28:48 & 16 & 14 & 0.1 & 0.45 & -- & -- & 0.7 & Y \\ 
OrionAS-536486-62838 & 5:36:48.6 & -6:28:38 & 51 & 33 & 0.3 & 2.4 & 1.3 & 13.1 & 3.8 & Y \\ 
OrionAS-535289-62751 & 5:35:28.9 & -6:27:51 & 31 & 19 & 0.2 & 0.7 & 0.6 & 2.0 & 1.1 & Y \\ 
OrionAS-535297-62701 & 5:35:29.7 & -6:27:01 & 27 & 19 & 0.9 & 1.7 & 2.0 & 4.9 & 2.7 & Y \\ 
OrionAS-535522-62634 & 5:35:52.2 & -6:26:34 & 22 & 16 & 0.1 & 0.3 & 0.5 & 2.2 & 0.5 & N \\ 
OrionAS-536318-62628 & 5:36:31.8 & -6:26:28 & 49 & 24 & 0.3 & 2.7 & 1.2 & 9.6 & 4.2 & N \\ 
OrionAS-536414-62627 & 5:36:41.4 & -6:26:27 & 61 & 37 & 0.7 & 5.8 & 2.3 & 24.9 & 9.1 & Y \\ 
OrionAS-535305-62622 & 5:35:30.5 & -6:26:22 & 32 & 21 & 0.4 & 1.6 & 1.2 & 5.2 & 2.5 & N \\ 
OrionAS-535474-62622 & 5:35:47.4 & -6:26:22 & 14 & 9 & 0.1 & 0.2 & 0.5 & 0.9 & 0.3 & Y \\ 
OrionAS-535566-62552 & 5:35:56.6 & -6:25:52 & 24 & 14 & 0.1 & 0.3 & 0.7 & 2.5 & 0.5 & N \\ 
OrionAS-536120-62552 & 5:36:12.0 & -6:25:52 & 21 & 15 & 0.1 & 0.09 & -- & -- & 0.1 & N \\ 
OrionAS-536367-62517 & 5:36:36.7 & -6:25:17 & 35 & 24 & 0.2 & 1.1 & 0.9 & 6.0 & 1.8 & N \\ 
OrionAS-536315-62516 & 5:36:31.5 & -6:25:16 & 25 & 21 & 0.3 & 1.2 & 1.2 & 5.6 & 1.9 & N \\ 
OrionAS-536288-62504 & 5:36:28.8 & -6:25:04 & 18 & 16 & 0.3 & 0.7 & 1.2 & 3.7 & 1.1 & N \\ 
OrionAS-536248-62454 & 5:36:24.8 & -6:24:54 & 43 & 38 & 1.0 & 7.1 & 3.4 & 29.5 & 11.1 & Y \\ 
OrionAS-535536-62428 & 5:35:53.6 & -6:24:28 & 20 & 12 & 0.1 & 0.4 & 0.7 & 2.6 & 0.6 & N \\ 
OrionAS-536170-62403 & 5:36:17.0 & -6:24:03 & 21 & 14 & 0.2 & 0.4 & 0.6 & 1.6 & 0.6 & N \\ 
OrionAS-536100-62354 & 5:36:10.0 & -6:23:54 & 18 & 12 & 0.1 & 0.2 & 0.4 & 0.7 & 0.2 & N \\ 
OrionAS-536184-62329 & 5:36:18.4 & -6:23:29 & 24 & 16 & 0.2 & 0.6 & 0.7 & 2.1 & 0.9 & N \\ 
OrionAS-536220-62325 & 5:36:22.0 & -6:23:25 & 19 & 17 & 0.3 & 0.6 & 0.9 & 2.7 & 1.0 & N \\ 
OrionAS-536250-62241 & 5:36:25.0 & -6:22:41 & 49 & 27 & 0.7 & 3.1 & 2.0 & 9.3 & 4.9 & Y \\ 
OrionAS-536196-62209 & 5:36:19.6 & -6:22:09 & 77 & 44 & 3.8 & 15.8 & 10.1 & 48.0 & 24.8 & Y \\ 
OrionAS-536286-61506 & 5:36:28.6 & -6:15:06 & 35 & 20 & 0.2 & 0.99 & -- & -- & 1.6 & N \\ 
OrionAS-536366-61459 & 5:36:36.6 & -6:14:59 & 19 & 13 & 0.4 & 0.76 & -- & -- & 1.2 & N \\ 
OrionAS-536247-61408 & 5:36:24.7 & -6:14:08 & 45 & 29 & 0.5 & 3.0 & 1.3 & 6.2 & 4.8 & N \\ 
OrionAS-535098-61358 & 5:35:09.8 & -6:13:58 & 57 & 39 & 0.4 & 4.5 & 1.7 & 23.9 & 7.0 & Y \\ 
OrionAS-535213-61313 & 5:35:21.3 & -6:13:13 & 26 & 17 & 0.2 & 0.7 & 0.9 & 3.8 & 1.0 & N \\ 
OrionAS-536204-61310 & 5:36:20.4 & -6:13:10 & 28 & 12 & 0.2 & 0.57 & -- & -- & 0.9 & N \\ 
OrionAS-536223-61258 & 5:36:22.3 & -6:12:58 & 20 & 16 & 0.3 & 0.57 & -- & -- & 0.9 & N \\ 
OrionAS-536364-61255 & 5:36:36.4 & -6:12:55 & 10 & 10 & 0.2 & 0.32 & -- & -- & 0.5 & N \\ 
OrionAS-535330-61245 & 5:35:33.0 & -6:12:45 & 17 & 14 & 0.1 & 0.2 & 0.6 & 1.8 & 0.3 & N \\ 
OrionAS-536176-61205 & 5:36:17.6 & -6:12:05 & 23 & 16 & 0.2 & 0.49 & -- & -- & 0.8 & N \\ 
OrionAS-536111-61044 & 5:36:11.1 & -6:10:44 & 56 & 19 & 0.7 & 3.2 & 2.1 & 11.3 & 5.0 & Y \\ 
OrionAS-535524-61007 & 5:35:52.4 & -6:10:07 & 33 & 19 & 0.5 & 1.3 & 1.4 & 4.7 & 2.1 & N \\ 
OrionAS-535373-60942 & 5:35:37.3 & -6:09:42 & 15 & 14 & 0.2 & 0.4 & 0.6 & 1.0 & 0.6 & N \\ 
OrionAS-535428-60754 & 5:35:42.8 & -6:07:54 & 51 & 33 & 0.3 & 3.3 & 1.1 & 14.1 & 5.2 & Y \\ 
OrionAS-535380-60723 & 5:35:38.0 & -6:07:23 & 29 & 17 & 0.2 & 0.7 & 0.9 & 3.9 & 1.1 & N \\ 
OrionAS-535405-60648 & 5:35:40.5 & -6:06:48 & 27 & 25 & 0.2 & 1.2 & 0.9 & 5.1 & 1.9 & N \\ 
OrionAS-535086-60602 & 5:35:08.6 & -6:06:02 & 22 & 14 & 0.2 & 0.43 & -- & -- & 0.7 & N \\ 
OrionAS-535361-60547 & 5:35:36.1 & -6:05:47 & 36 & 24 & 0.2 & 1.6 & 1.0 & 7.1 & 2.5 & Y \\ 
OrionAS-535405-60540 & 5:35:40.5 & -6:05:40 & 33 & 29 & 0.2 & 1.7 & 1.0 & 6.5 & 2.7 & N \\ 
OrionAS-535335-60500 & 5:35:33.5 & -6:05:00 & 35 & 24 & 0.2 & 1.5 & 1.0 & 6.4 & 2.4 & N \\ 
OrionAS-535083-60407 & 5:35:08.3 & -6:04:07 & 29 & 23 & 0.2 & 0.9 & 1.1 & 3.8 & 1.5 & N \\ 
OrionAS-535223-60316 & 5:35:22.3 & -6:03:16 & 35 & 23 & 0.2 & 1.5 & 0.7 & 4.1 & 2.4 & N \\ 
OrionAS-535127-60116 & 5:35:12.7 & -6:01:16 & 17 & 14 & 0.2 & 0.37 & -- & -- & 0.6 & N \\ 
OrionAS-535180-60010 & 5:35:18.0 & -6:00:10 & 37 & 15 & 0.2 & 1.2 & 0.7 & 3.5 & 2.0 & N \\ 
OrionAS-535274-55948 & 5:35:27.4 & -5:59:48 & 33 & 18 & 0.2 & 1.2 & 0.7 & 3.5 & 2.0 & N \\ 
OrionBN-546035-01453 & 5:46:03.5 & -0:14:53 & 18 & 12 & 0.6 & 0.9 & 1.0 & 1.4 & 1.4 & Y \\ 
OrionBN-546051-01419 & 5:46:05.1 & -0:14:19 & 20 & 16 & 0.4 & 0.9 & 0.8 & 1.3 & 1.4 & Y \\ 
\hline
\end{tabular}
\end{center}
\end{table*}
\begin{table*}
\begin{center}
\contcaption{Core properties}
\begin{tabular}{|l|c|c|c|c|c|c|c|c|c|c|}\hline
&&& \multicolumn{2}{c|}{Aperture} & \multicolumn{2}{c|}{850~$\mu$m} & \multicolumn{2}{c|}{450~$\mu$m} & & \\ 
Source 	& RA 		& Dec.		& Semi-major& Semi-minor  & Peak	&Int.	& Peak		& Int. & Mass & YSO \\
Name	&(2000)		&(2000)		& (arcsec)  &(arcsec) &(Jy/beam)	&(Jy)	&(Jy/beam)	&(Jy)  & (M${_\odot}$) & \\ \hline
OrionBN-546074-01342 & 5:46:07.4 & -0:13:42 & 49 & 23 & 1.6 & 4.0 & 4.9 & 10.8 & 6.3 & N \\ 
OrionBN-546234-01323 & 5:46:23.4 & -0:13:23 & 21 & 19 & 0.2 & 0.5 & 0.4 & 1.1 & 0.8 & N \\ 
OrionBN-546078-01223 & 5:46:07.8 & -0:12:23 & 15 & 13 & 0.4 & 0.5 & 0.8 & 1.1 & 0.8 & N \\ 
OrionBN-546103-01219 & 5:46:10.3 & -0:12:19 & 19 & 17 & 0.6 & 0.8 & 1.0 & 1.4 & 1.3 & N \\ 
OrionBN-546046-01216 & 5:46:04.6 & -0:12:16 & 32 & 15 & 0.3 & 0.7 & 0.6 & 0.8 & 1.1 & Y \\ 
OrionBN-546079-01150 & 5:46:07.9 & -0:11:50 & 21 & 18 & 0.7 & 1.1 & 1.5 & 2.8 & 1.8 & N \\ 
OrionBN-546086-01048 & 5:46:08.6 & -0:10:48 & 25 & 22 & 3.0 & 3.9 & 4.3 & 7.2 & 6.1 & Y \\ 
OrionBN-546138-01008 & 5:46:13.8 & -0:10:08 & 15 & 10 & 0.2 & 0.2 & 0.4 & 0.7 & 0.4 & N \\ 
OrionBN-546080-01002 & 5:46:08.0 & -0:10:02 & 33 & 19 & 0.7 & 1.9 & 2.0 & 5.9 & 2.9 & Y \\ 
OrionBN-546126-00946 & 5:46:12.6 & -0:09:46 & 29 & 15 & 0.2 & 0.6 & 0.6 & 1.6 & 1.0 & N \\ 
OrionBN-546063-00935 & 5:46:06.3 & -0:09:35 & 24 & 18 & 0.4 & 1.1 & 1.0 & 3.0 & 1.7 & N \\ 
OrionBN-546091-00922 & 5:46:09.1 & -0:09:22 & 19 & 18 & 0.2 & 0.5 & -- & -- & 0.7 & N \\ 
OrionBN-546115-00917 & 5:46:11.5 & -0:09:17 & 14 & 9 & 0.2 & 0.2 & -- & -- & 0.3 & N \\ 
OrionBN-546049-00911 & 5:46:04.9 & -0:09:11 & 15 & 12 & 0.2 & 0.3 & 0.5 & 1.0 & 0.4 & N \\ 
OrionBN-546131-00606 & 5:46:13.1 & -0:06:06 & 16 & 13 & 0.2 & 0.3 & 0.4 & 0.6 & 0.5 & Y \\ 
OrionBN-546097-00552 & 5:46:09.7 & -0:05:52 & 23 & 13 & 0.1 & 0.3 & 0.2 & 0.4 & 0.5 & N \\ 
OrionBN-546135-00525 & 5:46:13.5 & -0:05:25 & 24 & 20 & 0.2 & 0.7 & 0.5 & 1.0 & 1.1 & N \\ 
OrionBN-546195-00523 & 5:46:19.5 & -0:05:23 & 12 & 9 & 0.1 & 0.1 & -- & -- & 0.2 & Y \\ 
OrionBN-547031-00239 & 5:47:03.1 & -0:02:39 & 28 & 10 & 0.2 & 0.3 & -- & -- & 0.5 & N \\ 
OrionBN-546310-00234 & 5:46:31.0 & -0:02:34 & 28 & 24 & 1.5 & 2.7 & 3.7 & 5.6 & 4.2 & N \\ 
OrionBN-546433-00148 & 5:46:43.3 & -0:01:48 & 34 & 17 & 0.2 & 0.5 & 0.3 & 0.4 & 0.8 & N \\ 
OrionBN-546280-00145 & 5:46:28.0 & -0:01:45 & 21 & 21 & 0.4 & 1.2 & 0.7 & 1.8 & 1.9 & N \\ 
OrionBN-546509-00129 & 5:46:50.9 & -0:01:29 & 24 & 17 & 0.2 & 0.5 & -- & -- & 0.8 & Y \\ 
OrionBN-546276-00057 & 5:46:27.6 & -0:00:57 & 25 & 21 & 1.2 & 2.5 & 2.1 & 4.3 & 3.9 & N \\ 
OrionBN-546321-00044 & 5:46:32.1 & -0:00:44 & 20 & 15 & 0.3 & 0.6 & 0.6 & 1.0 & 0.9 & N \\ 
OrionBN-546532-00018 & 5:46:53.2 & -0:00:18 & 66 & 18 & 0.3 & 3.1 & 0.9 & 7.4 & 4.8 & N \\ 
OrionBN-546334-00006 & 5:46:33.4 & -0:00:06 & 22 & 16 & 0.3 & 0.7 & 0.8 & 1.5 & 1.1 & N \\ 
OrionBN-546244-00001 & 5:46:24.4 & -0:00:01 & 35 & 24 & 0.4 & 1.8 & 1.1 & 4.6 & 2.8 & N \\ 
OrionBN-546450+00021 & 5:46:45.0 & 0:00:21 & 18 & 17 & 0.4 & 1.1 & 1.0 & 3.0 & 1.7 & N \\ 
OrionBN-546474+00027 & 5:46:47.4 & 0:00:27 & 27 & 21 & 1.2 & 3.0 & 2.7 & 6.8 & 4.8 & Y \\ 
OrionBN-546350+00029 & 5:46:35.0 & 0:00:29 & 26 & 18 & 0.4 & 1.0 & 1.0 & 2.8 & 1.6 & N \\ 
OrionBN-546405+00032 & 5:46:40.5 & 0:00:32 & 21 & 20 & 0.5 & 1.7 & 1.2 & 4.4 & 2.7 & N \\ 
OrionBN-546378+00035 & 5:46:37.8 & 0:00:35 & 21 & 17 & 0.7 & 1.4 & 1.6 & 3.5 & 2.2 & Y \\ 
OrionBN-546430+00047 & 5:46:43.0 & 0:00:47 & 26 & 25 & 0.7 & 2.8 & 1.8 & 6.9 & 4.4 & Y \\ 
OrionBN-546267+00101 & 5:46:26.7 & 0:01:01 & 26 & 19 & 0.5 & 1.2 & 1.2 & 3.1 & 1.8 & N \\ 
OrionBN-546397+00109 & 5:46:39.7 & 0:01:09 & 22 & 21 & 0.8 & 1.9 & 1.8 & 3.8 & 2.9 & Y \\ 
OrionBN-546490+00116 & 5:46:49.0 & 0:01:16 & 50 & 25 & 0.7 & 5.2 & 1.9 & 12.7 & 8.1 & Y \\ 
OrionBN-546275+00135 & 5:46:27.5 & 0:01:35 & 22 & 16 & 0.3 & 0.9 & 0.9 & 2.3 & 1.4 & N \\ 
OrionBN-546499+00204 & 5:46:49.9 & 0:02:04 & 49 & 23 & 0.8 & 4.3 & 2.2 & 11.9 & 6.7 & N \\ 
OrionBN-546398+00420 & 5:46:39.8 & 0:04:20 & 20 & 20 & 0.2 & 0.6 & 0.3 & 0.7 & 0.9 & N \\ 
OrionBN-546288+00435 & 5:46:28.8 & 0:04:35 & 29 & 12 & 0.1 & 0.4 & 0.4 & 1.2 & 0.6 & N \\ 
OrionBN-546544+00440 & 5:46:54.4 & 0:04:40 & 27 & 14 & 0.1 & 0.3 & 0.2 & 0.4 & 0.4 & N \\ 
OrionBN-546335+00526 & 5:46:33.5 & 0:05:26 & 26 & 12 & 0.2 & 0.4 & 0.3 & 0.7 & 0.6 & N \\ 
OrionBN-546362+00550 & 5:46:36.2 & 0:05:50 & 20 & 15 & 0.3 & 0.6 & 0.5 & 1.0 & 0.9 & N \\ 
OrionBN-546462+00717 & 5:46:46.2 & 0:07:17 & 44 & 40 & 0.5 & 4.4 & 1.2 & 11.1 & 6.9 & Y \\ 
OrionBN-546496+00902 & 5:46:49.6 & 0:09:02 & 14 & 14 & 0.2 & 0.3 & 0.4 & 0.8 & 0.5 & N \\ 
OrionBN-546516+00916 & 5:46:51.6 & 0:09:16 & 24 & 18 & 0.2 & 0.7 & 0.6 & 1.8 & 1.0 & N \\ 
OrionBN-546484+01012 & 5:46:48.4 & 0:10:12 & 15 & 13 & 0.1 & 0.3 & 0.4 & 0.8 & 0.4 & N \\ 
OrionBN-547237+01102 & 5:47:23.7 & 0:11:02 & 26 & 15 & 0.2 & 0.4 & -- & -- & 0.6 & N \\ 
OrionBN-547068+01230 & 5:47:06.8 & 0:12:30 & 27 & 22 & 0.4 & 1.2 & 0.9 & 2.7 & 1.9 & N \\ 
OrionBN-547106+01318 & 5:47:10.6 & 0:13:18 & 32 & 18 & 0.2 & 0.6 & 0.5 & 1.7 & 1.0 & N \\ 
OrionBN-547051+01321 & 5:47:05.1 & 0:13:21 & 23 & 13 & 0.2 & 0.3 & 0.4 & 0.7 & 0.5 & N \\ 
OrionBN-547050+01449 & 5:47:05.0 & 0:14:49 & 38 & 26 & 0.3 & 1.5 & 0.7 & 4.6 & 2.4 & N \\ 
OrionBN-547239+01507 & 5:47:23.9 & 0:15:07 & 64 & 23 & 0.2 & 1.9 & 0.6 & 6.1 & 3.0 & N \\ 
OrionBN-547124+01537 & 5:47:12.4 & 0:15:37 & 16 & 13 & 0.2 & 0.3 & 0.7 & 1.4 & 0.5 & N \\ 
OrionBN-547104+01553 & 5:47:10.4 & 0:15:53 & 16 & 14 & 0.2 & 0.3 & 0.6 & 1.4 & 0.5 & N \\ 
OrionBN-547199+01603 & 5:47:19.9 & 0:16:03 & 23 & 16 & 0.2 & 0.6 & 0.6 & 1.6 & 0.9 & Y \\ 
OrionBN-547048+01707 & 5:47:04.8 & 0:17:07 & 21 & 21 & 0.2 & 0.7 & 0.7 & 2.4 & 1.1 & N \\ 
OrionBN-547015+01755 & 5:47:01.5 & 0:17:55 & 35 & 30 & 0.4 & 2.2 & 1.1 & 7.0 & 3.5 & Y \\ 
OrionBN-547087+01817 & 5:47:08.7 & 0:18:17 & 15 & 11 & 0.1 & 0.3 & 0.6 & 1.1 & 0.4 & N \\ 
\hline
\end{tabular}
\end{center}
\end{table*}
\begin{table*}
\begin{center}
\contcaption{Core properties}
\begin{tabular}{|l|c|c|c|c|c|c|c|c|c|c|}\hline
&&& \multicolumn{2}{c|}{Aperture} & \multicolumn{2}{c|}{850~$\mu$m} & \multicolumn{2}{c|}{450~$\mu$m} & & \\ 
Source 	& RA 		& Dec.		& Semi-major& Semi-minor  & Peak	&Int.	& Peak		& Int. & Mass & YSO \\
Name	&(2000)		&(2000)		& (arcsec)  &(arcsec) &(Jy/beam)	&(Jy)	&(Jy/beam)	&(Jy)  & (M${_\odot}$) & \\ \hline
OrionBN-547152+01830 & 5:47:15.2 & 0:18:30 & 56 & 28 & 0.3 & 2.4 & 0.7 & 5.8 & 3.7 & Y \\ 
OrionBN-547253+01848 & 5:47:25.3 & 0:18:48 & 21 & 12 & 0.2 & 0.4 & 0.4 & 0.6 & 0.6 & N \\ 
OrionBN-547336+01902 & 5:47:33.6 & 0:19:02 & 16 & 9 & 0.2 & 0.2 & -- & -- & 0.4 & N \\ 
OrionBN-546283+01928 & 5:46:28.3 & 0:19:28 & 24 & 19 & 1.1 & 1.7 & 1.3 & 2.6 & 2.7 & Y \\ 
OrionBN-547034+01950 & 5:47:03.4 & 0:19:50 & 62 & 17 & 0.3 & 0.2 & 0.8 & 3.3 & 0.4 & N \\ 
OrionBN-547267+01953 & 5:47:26.7 & 0:19:53 & 39 & 16 & 0.3 & 1.2 & 0.5 & 1.9 & 1.9 & N \\ 
OrionBN-547377+02001 & 5:47:37.7 & 0:20:01 & 31 & 21 & 0.6 & 1.9 & 1.3 & 4.3 & 2.9 & Y \\ 
OrionBN-546576+02009 & 5:46:57.6 & 0:20:09 & 26 & 17 & 0.4 & 0.9 & -- & -- & 1.4 & N \\ 
OrionBN-546294+02010 & 5:46:29.4 & 0:20:10 & 21 & 20 & 0.4 & 1.2 & 0.6 & 2.2 & 1.9 & N \\ 
OrionBN-547349+02020 & 5:47:34.9 & 0:20:20 & 17 & 16 & 0.3 & 0.6 & 0.6 & 1.5 & 1.0 & N \\ 
OrionBN-547325+02026 & 5:47:32.5 & 0:20:26 & 21 & 17 & 0.5 & 0.9 & 0.9 & 1.6 & 1.4 & Y \\ 
OrionBN-547010+02042 & 5:47:01.0 & 0:20:42 & 29 & 19 & 0.3 & 1.0 & -- & -- & 1.6 & Y \\ 
OrionBN-547252+02059 & 5:47:25.2 & 0:20:59 & 39 & 25 & 1.5 & 4.4 & 2.6 & 10.0 & 6.9 & N \\ 
OrionBN-547103+02112 & 5:47:10.3 & 0:21:12 & 24 & 20 & 1.1 & 1.9 & 2.0 & 5.4 & 3.0 & N \\ 
OrionBN-546287+02114 & 5:46:28.7 & 0:21:14 & 42 & 21 & 0.2 & 0.7 & 0.2 & 1.9 & 1.1 & N \\ 
OrionBN-547160+02123 & 5:47:16.0 & 0:21:23 & 32 & 15 & 0.4 & 1.0 & 1.0 & 3.3 & 1.6 & N \\ 
OrionBN-547041+02158 & 5:47:04.1 & 0:21:58 & 68 & 36 & 9.0 & 31.8 & 16.7 & 71.6 & 50.0 & Y \\ 
OrionBN-546253+02220 & 5:46:25.3 & 0:22:20 & 39 & 20 & 0.1 & 0.5 & 0.2 & 1.7 & 0.8 & N \\ 
OrionBN-546528+02223 & 5:46:52.8 & 0:22:23 & 24 & 14 & 0.2 & 0.5 & 0.6 & 1.9 & 0.9 & N \\ 
OrionBN-547119+02223 & 5:47:11.9 & 0:22:23 & 26 & 13 & 0.4 & 0.8 & 0.7 & 1.9 & 1.2 & Y \\ 
OrionBN-547197+02231 & 5:47:19.7 & 0:22:31 & 16 & 10 & 0.2 & 0.3 & -- & -- & 0.5 & N \\ 
OrionBN-547069+02239 & 5:47:06.9 & 0:22:39 & 32 & 18 & 0.7 & 2.9 & 0.8 & 3.3 & 4.5 & Y \\ 
OrionBN-547175+02240 & 5:47:17.5 & 0:22:40 & 17 & 11 & 0.2 & 0.3 & -- & -- & 0.5 & N \\ 
OrionBN-546591+02259 & 5:46:59.1 & 0:22:59 & 31 & 16 & 0.3 & 1.1 & 0.8 & 2.9 & 1.7 & N \\ 
OrionBN-547124+02311 & 5:47:12.4 & 0:23:11 & 30 & 16 & 0.3 & 0.9 & 0.6 & 1.7 & 1.4 & N \\ 
OrionBN-547067+02314 & 5:47:06.7 & 0:23:14 & 16 & 11 & 0.2 & 0.4 & -- & -- & 0.6 & N \\ 
OrionBN-546547+02324 & 5:46:54.7 & 0:23:24 & 46 & 26 & 0.4 & 3.5 & 0.8 & 6.8 & 5.4 & Y \\ 
OrionBN-547104+02327 & 5:47:10.4 & 0:23:27 & 25 & 15 & 0.3 & 0.8 & -- & -- & 1.3 & N \\ 
OrionBN-547089+02356 & 5:47:08.9 & 0:23:56 & 24 & 14 & 0.1 & 0.2 & -- & -- & 0.3 & N \\ 
OrionBN-546572+02356 & 5:46:57.2 & 0:23:56 & 20 & 16 & 0.4 & 1.0 & 0.7 & 1.9 & 1.5 & Y \\ 
OrionBN-546347+02359 & 5:46:34.7 & 0:23:59 & 52 & 27 & 0.2 & 1.3 & 0.3 & 1.5 & 2.1 & N \\ 
OrionBN-546580+02426 & 5:46:58.0 & 0:24:26 & 25 & 14 & 0.2 & 0.6 & 0.5 & 1.6 & 1.0 & N \\ 
OrionBN-547017+02452 & 5:47:01.7 & 0:24:52 & 21 & 11 & 0.1 & 0.1 & -- & -- & 0.2 & Y \\ 
OrionBN-546257+02456 & 5:46:25.7 & 0:24:56 & 29 & 22 & 0.1 & 0.5 & -- & -- & 0.8 & N \\ 
OrionBN-547080+02505 & 5:47:08.0 & 0:25:05 & 10 & 9 & 0.1 & 0.2 & -- & -- & 0.2 & N \\ 
OrionBN-546459+02507 & 5:46:45.9 & 0:25:07 & 47 & 47 & 0.1 & 1.8 & -- & -- & 2.9 & N \\ 
OrionBN-547014+02614 & 5:47:01.4 & 0:26:14 & 33 & 21 & 0.8 & 1.8 & 1.0 & 1.8 & 2.9 & N \\ 
OrionBN-546380+02653 & 5:46:38.0 & 0:26:53 & 26 & 19 & 0.2 & 0.8 & -- & -- & 1.2 & N \\ 
OrionBS-541356-23011 & 5:41:35.6 & -2:30:11 & 27 & 13 & 0.2 & 0.6 & 0.6 & 2.2 & 1.0 & N \\ 
OrionBS-541236-22901 & 5:41:23.6 & -2:29:01 & 27 & 15 & 0.2 & 0.5 & 0.5 & 1.3 & 0.8 & Y \\ 
OrionBS-541157-22830 & 5:41:15.7 & -2:28:30 & 20 & 18 & 0.1 & 0.4 & 0.3 & 1.4 & 0.6 & N \\ 
OrionBS-541195-22752 & 5:41:19.5 & -2:27:52 & 29 & 15 & 0.1 & 0.5 & 0.2 & 1.1 & 0.8 & N \\ 
OrionBS-541321-22751 & 5:41:32.1 & -2:27:51 & 17 & 10 & 0.1 & 0.2 & -- & -- & 0.4 & N \\ 
OrionBS-540544-22733 & 5:40:54.4 & -2:27:33 & 80 & 27 & 0.3 & 1.5 & 0.7 & 2.8 & 2.3 & Y \\ 
OrionBS-541070-22720 & 5:41:07.0 & -2:27:20 & 47 & 19 & 0.4 & 1.6 & 1.1 & 5.7 & 2.5 & N \\ 
OrionBS-540584-22716 & 5:40:58.4 & -2:27:16 & 15 & 11 & 0.1 & 0.1 & -- & -- & 0.2 & N \\ 
OrionBS-541315-22542 & 5:41:31.5 & -2:25:42 & 64 & 36 & 0.2 & 2.9 & 0.8 & 12.2 & 4.6 & Y \\ 
OrionBS-540579-22534 & 5:40:57.9 & -2:25:34 & 21 & 16 & 0.1 & 0.1 & 0.2 & 0.6 & 0.2 & N \\ 
OrionBS-541280-22409 & 5:41:28.0 & -2:24:09 & 38 & 27 & 0.2 & 1.2 & 0.8 & 6.3 & 1.9 & N \\ 
OrionBS-541294-22318 & 5:41:29.4 & -2:23:18 & 25 & 14 & 0.2 & 0.5 & -- & -- & 0.8 & N \\ 
OrionBS-541299-22114 & 5:41:29.9 & -2:21:14 & 30 & 21 & 0.8 & 2.1 & 1.9 & 6.7 & 3.3 & N \\ 
OrionBS-541334-22100 & 5:41:33.4 & -2:21:00 & 18 & 13 & 0.2 & 0.4 & 0.9 & 2.1 & 0.7 & N \\ 
OrionBS-541193-22056 & 5:41:19.3 & -2:20:56 & 16 & 8 & 0.1 & 0.2 & -- & -- & 0.3 & N \\ 
OrionBS-541289-22005 & 5:41:28.9 & -2:20:05 & 38 & 20 & 0.4 & 1.7 & 1.0 & 4.3 & 2.7 & N \\ 
OrionBS-541471-21930 & 5:41:47.1 & -2:19:30 & 25 & 20 & 0.2 & 0.6 & 0.9 & 4.8 & 1.0 & N \\ 
OrionBS-541258-21925 & 5:41:25.8 & -2:19:25 & 30 & 21 & 0.4 & 1.4 & 1.1 & 3.3 & 2.2 & N \\ 
OrionBS-541312-21902 & 5:41:31.2 & -2:19:02 & 25 & 15 & 0.2 & 0.7 & -- & -- & 1.1 & N \\ 
OrionBS-541257-21809 & 5:41:25.7 & -2:18:09 & 42 & 30 & 2.8 & 8.1 & 4.7 & 17.5 & 12.7 & Y \\ 
OrionBS-541375-21733 & 5:41:37.5 & -2:17:33 & 47 & 30 & 1.2 & 5.3 & 1.8 & 16.4 & 8.4 & Y \\ 
OrionBS-541241-21711 & 5:41:24.1 & -2:17:11 & 27 & 23 & 0.4 & 1.8 & 0.9 & 4.0 & 2.8 & N \\ 
\hline
\end{tabular}
\end{center}
\end{table*}
\begin{table*}
\begin{center}
\contcaption{Core properties}
\begin{tabular}{|l|c|c|c|c|c|c|c|c|c|c|}\hline
&&& \multicolumn{2}{c|}{Aperture} & \multicolumn{2}{c|}{850~$\mu$m} & \multicolumn{2}{c|}{450~$\mu$m} & & \\ 
Source 	& RA 		& Dec.		& Semi-major& Semi-minor  & Peak	&Int.	& Peak		& Int. & Mass & YSO \\
Name	&(2000)		&(2000)		& (arcsec)  &(arcsec) &(Jy/beam)	&(Jy)	&(Jy/beam)	&(Jy)  & (M${_\odot}$) & \\ \hline
OrionBS-541446-21704 & 5:41:44.6 & -2:17:04 & 31 & 28 & 0.3 & 1.6 & 1.1 & 6.9 & 2.4 & Y \\ 
OrionBS-541404-21702 & 5:41:40.4 & -2:17:02 & 51 & 28 & 0.8 & 5.2 & 2.4 & 18.1 & 8.1 & Y \\ 
OrionBS-541429-21603 & 5:41:42.9 & -2:16:03 & 40 & 28 & 0.4 & 2.1 & 1.3 & 7.0 & 3.3 & Y \\ 
OrionBS-541251-21602 & 5:41:25.1 & -2:16:02 & 39 & 29 & 0.4 & 3.0 & 1.0 & 7.5 & 4.7 & Y \\ 
OrionBS-541161-21534 & 5:41:16.1 & -2:15:34 & 33 & 17 & 0.2 & 0.7 & -- & -- & 1.1 & N \\ 
OrionBS-540590-20853 & 5:40:59.0 & -2:08:53 & 66 & 30 & 0.3 & 3.3 & 0.9 & 13.1 & 5.1 & N \\ 
OrionBS-542480-20839 & 5:42:48.0 & -2:08:39 & 52 & 16 & 0.2 & 1.0 & -- & -- & 1.5 & N \\ 
OrionBS-541542-20811 & 5:41:54.2 & -2:08:11 & 25 & 16 & 0.1 & 0.3 & -- & -- & 0.4 & Y \\ 
OrionBS-542025-20739 & 5:42:02.5 & -2:07:39 & 32 & 20 & 0.9 & 1.6 & 1.5 & 3.3 & 2.5 & N \\ 
OrionBS-540575-20728 & 5:40:57.5 & -2:07:28 & 18 & 16 & 0.2 & 0.4 & -- & -- & 0.7 & Y \\ 
OrionBS-542030-20423 & 5:42:03.0 & -2:04:23 & 23 & 13 & 0.2 & 0.4 & 0.4 & 1.3 & 0.6 & N \\ 
OrionBS-542103-20420 & 5:42:10.3 & -2:04:20 & 22 & 13 & 0.3 & 0.6 & 0.8 & 1.5 & 0.9 & N \\ 
OrionBS-542035-20224 & 5:42:03.5 & -2:02:24 & 45 & 25 & 1.0 & 3.0 & 2.8 & 8.0 & 4.7 & N \\ 
OrionBS-541571-20100 & 5:41:57.1 & -2:01:00 & 20 & 17 & 0.3 & 0.8 & 0.9 & 2.9 & 1.2 & N \\ 
OrionBS-541529-20021 & 5:41:52.9 & -2:00:21 & 51 & 23 & 0.3 & 1.9 & 1.2 & 9.4 & 3.0 & Y \\ 
OrionBS-541493-15938 & 5:41:49.3 & -1:59:38 & 29 & 15 & 0.4 & 1.2 & 1.6 & 5.1 & 1.9 & Y \\ 
OrionBS-541107-15809 & 5:41:10.7 & -1:58:09 & 28 & 14 & 0.1 & 0.3 & 0.5 & 0.6 & 0.4 & N \\ 
OrionBS-542000-15801 & 5:42:00.0 & -1:58:01 & 18 & 16 & 0.1 & 0.3 & -- & -- & 0.5 & N \\ 
OrionBS-541491-15803 & 5:41:49.1 & -1:58:03 & 63 & 32 & 0.9 & 7.5 & 3.6 & 34.9 & 11.9 & Y \\ 
OrionBS-541452-15631 & 5:41:45.2 & -1:56:31 & 44 & 25 & 8.1 & 21.7 & 22.4 & 70.4 & 58.7 & Y \\ 
OrionBS-541354-15629 & 5:41:35.4 & -1:56:29 & 48 & 30 & 0.3 & 2.2 & 0.5 & 2.0 & 3.5 & N \\ 
OrionBS-541031-15554 & 5:41:03.1 & -1:55:54 & 14 & 13 & 0.2 & 0.2 & 0.5 & 1.3 & 0.4 & N \\ 
OrionBS-541445-15539 & 5:41:44.5 & -1:55:39 & 42 & 22 & 12.4 & 42.5 & 30.7 & 104.4 & 104.5 & N \\ 
OrionBS-541442-15443 & 5:41:44.2 & -1:54:43 & 24 & 19 & 6.3 & 16.5 & 20.4 & 49.8 & 45.2 & N \\ 
OrionBS-541321-15426 & 5:41:32.1 & -1:54:26 & 20 & 11 & 0.2 & 0.3 & 0.4 & 0.9 & 0.5 & N \\ 
OrionBS-541199-15416 & 5:41:19.9 & -1:54:16 & 15 & 10 & 0.1 & 0.2 & -- & -- & 0.2 & N \\ 
OrionBS-541420-15359 & 5:41:42.0 & -1:53:59 & 42 & 27 & 9.3 & 31.7 & 28.7 & 92.6 & 78.3 & N \\ 
OrionBS-541362-15256 & 5:41:36.2 & -1:52:56 & 18 & 14 & 0.2 & 0.4 & 0.5 & 1.4 & 0.5 & Y \\ 
OrionBS-541442-15241 & 5:41:44.2 & -1:52:41 & 28 & 21 & 0.6 & 2.8 & 2.1 & 8.2 & 4.4 & Y \\ 
OrionBS-541166-15119 & 5:41:16.6 & -1:51:19 & 30 & 12 & 0.2 & 0.5 & 0.7 & 2.3 & 0.8 & N \\ 
OrionBS-541367-15106 & 5:41:36.7 & -1:51:06 & 82 & 40 & 0.4 & 8.2 & 1.0 & 12.3 & 12.9 & Y \\ 
OrionBS-541234-15027 & 5:41:23.4 & -1:50:27 & 32 & 20 & 0.3 & 1.1 & 1.1 & 3.5 & 1.7 & N \\ 
OrionBS-541329-14953 & 5:41:32.9 & -1:49:53 & 43 & 28 & 0.4 & 2.3 & 0.5 & 0.4 & 3.7 & N \\ 
OrionBS-541364-14924 & 5:41:36.4 & -1:49:24 & 49 & 23 & 0.5 & 2.8 & 1.0 & 1.8 & 4.4 & Y \\ 
OrionBS-541276-14813 & 5:41:27.6 & -1:48:13 & 27 & 11 & 0.2 & 0.4 & -- & -- & 0.7 & Y \\ 
OrionBS-541113-14812 & 5:41:11.3 & -1:48:12 & 33 & 21 & 0.6 & 2.6 & 2.0 & 9.2 & 4.1 & N \\ 
OrionBS-541133-14735 & 5:41:13.3 & -1:47:35 & 33 & 18 & 0.6 & 2.3 & 2.1 & 8.4 & 3.6 & Y \\ 
\hline
\end{tabular}
\end{center}
\end{table*}
\end{document}